\documentclass[aps,pra,twocolumn,floatfix,longbibliography, superscriptaddress]{revtex4-1}

\usepackage{pifont}
\usepackage[dvipsnames]{xcolor}
\definecolor{bananayellow}{rgb}{1.0, 0.88, 0.21}
\definecolor{amethyst}{rgb}{0.6, 0.4, 0.8}
\definecolor{ao(english)}{rgb}{0.0, 0.5, 0.0}

\usepackage{float}
\usepackage{epsfig}
\usepackage{bm}
\usepackage[matrix,frame,arrow]{xy}
\usepackage[applemac]{inputenc}
\usepackage[T1]{fontenc}
\usepackage{lmodern}
\usepackage[english]{babel}
\usepackage{amsmath}
\usepackage{ae}
\usepackage{units}
\usepackage{amssymb}
\usepackage{color}
\usepackage{graphicx}
\usepackage{bbm}
\usepackage{url}
\usepackage{nomencl}
\usepackage{subfigure}
\usepackage{slashed}

\usepackage{tikz}
\usetikzlibrary{arrows,snakes,calc}

\usepackage{fixmath}
\usepackage{booktabs}

\newcommand{\bra}[1]{\langle #1|}
\newcommand{\ket}[1]{|#1\rangle}

\newcommand{\sz}{\hat \sigma_z}
\newcommand{\sx}{\hat \sigma_x}
\newcommand{\sm}{\hat \sigma_-}
\renewcommand{\sp}{\hat \sigma_+}

\newcommand{\nn}{\nonumber}

\newcommand{\figref}[1]{\mbox{Fig.~\ref{#1}}}

\newcommand{\appref}[1]{\mbox{Appendix \ref{#1}}}
\renewcommand{\eqref}[1]{\mbox{Eq.~(\ref{#1})}}

\newcommand{\be}{\begin{equation}}
\newcommand{\ee}{\end{equation}}
\newcommand{\bea}{\begin{eqnarray}}
\newcommand{\eea}{\end{eqnarray}}

\usepackage{xr}
\usepackage[colorlinks]{hyperref}
\hypersetup{%
    plainpages=true,
    breaklinks=true,
    hypertexnames=false,
    pageanchor=true,
    colorlinks=true,
    linkcolor={blue},
    citecolor={red},
    urlcolor={blue},
    anchorcolor={black}
}

\newcommand{\beq}{\begin{eqnarray}}
\newcommand{\eeq}{\end{eqnarray}}

\begin{document}


\title{Quantum nonlinear optics without photons}

\author{Roberto Stassi}
\affiliation{Center for Emergent Matter Science, RIKEN, Saitama
351-0198, Japan}
\author{Vincenzo Macr\`{i}}
\affiliation{Dipartimento di Scienze Matematiche e Informatiche,
Scienze Fisiche e  Scienze della Terra,
    Universit\`{a} di Messina, I-98166 Messina, Italy}
\affiliation{Center for Emergent Matter Science, RIKEN, Saitama
    351-0198, Japan}
\author{Anton Frisk Kockum}
\affiliation{Center for Emergent Matter Science, RIKEN, Saitama
351-0198, Japan}
\author{Omar Di Stefano}
\affiliation{Dipartimento di Scienze Matematiche e Informatiche,
Scienze Fisiche e  Scienze della Terra,
    Universit\`{a} di Messina, I-98166 Messina, Italy}
\affiliation{Center for Emergent Matter Science, RIKEN, Saitama
351-0198, Japan}
\author{Adam Miranowicz}
\affiliation{Faculty of Physics, Adam Mickiewicz University,
PL-61-614 Poznan, Poland} \affiliation{Center for Emergent Matter
Science, RIKEN, Saitama 351-0198, Japan}
\author{Salvatore Savasta}
\affiliation{Dipartimento di Scienze Matematiche e Informatiche,
Scienze Fisiche e  Scienze della Terra, Universit\`{a} di Messina,
I-98166 Messina, Italy} \affiliation{Center for Emergent Matter
Science, RIKEN, Saitama 351-0198, Japan}
\author{Franco Nori}
\affiliation{Center for Emergent Matter Science, RIKEN, Saitama
351-0198, Japan} \affiliation{Physics Department, The University
of Michigan, Ann Arbor, Michigan 48109-1040, USA}

\begin{abstract}
Spontaneous parametric down-conversion is a well-known process in
quantum nonlinear optics in which a photon incident on a nonlinear
crystal spontaneously splits into two  photons. Here we propose an
analogous physical process where one excited atom directly
transfers its excitation to a pair of spatially-separated atoms
with probability approaching one. The interaction is mediated by
the exchange of {\it virtual} rather than {\it real} photons. This
nonlinear atomic process is coherent and reversible, so the pair
of excited atoms can transfer  the excitation back to the first
one: the atomic analog of sum-frequency generation of light. The
parameters used to investigate this process correspond to
experimentally-demonstrated values in ultrastrong circuit quantum
electrodynamics. This approach can be extended to realize other
nonlinear interatomic processes, such as four-atom mixing, and is
an attractive architecture for the realization of quantum devices
on a chip. We show that four-qubit mixing can efficiently
implement quantum repetition codes and, thus, can be used for
error-correction codes.


\end{abstract}


\maketitle


\section{Introduction}

It is highly desirable to couple distant qubits for
quantum-information applications~\cite{Duan2001,Chou2007}. One
implementation of such a quantum bus has been demonstrated using
microwave photons confined in a transmission line cavity, to
couple two superconducting qubits on opposite sides of a chip
\cite{Majer2007}. Interestingly, the interaction between the two
qubits is mediated by the exchange of {\em virtual} rather than
{\em real} photons, avoiding cavity-induced
losses~\cite{Majer2007}. The effective qubit-qubit interaction is
the result of the exchange of virtual photons with the cavity, and
gives rise to a qubit-qubit avoided-level crossing. At the minimum
splitting, the eigenstates of the system are symmetric and
antisymmetric superpositions of the two qubit states $|e,g
\rangle$ and $|g,e \rangle$, where $|g \rangle$ ($|e \rangle$)
labels the ground (excited)  state of the qubits. When the two
qubits have the {\it same} transition energy, an excitation in one
qubit can be coherently transferred to the other qubit by
virtually becoming a photon in the cavity~\cite{Majer2007}. When
the qubits have different transition energies, the interaction is
effectively turned off. The absence of cavity-induced losses, due
to the virtual nature of the quantum bus, is useful especially in
the presence of intrinsically-lossy interaction channels. For
example, it has been shown
theoretically~\cite{Gonzalez-Tudela2011} that virtual plasmon
polaritons in realistic one-dimensional subwavelength  waveguides
are excellent candidates to act as mediators for achieving a high
degree of entanglement between two distant qubits.

Here we propose a generalization of the qubit-qubit coupling via
virtual bosons, which enables the interaction of multiple ($N >2$)
spatially-separated qubits with {\it different} transition
frequencies. For example, we show that, in analogy to the
frequency-mixing processes of nonlinear optics, one qubit of given
transition frequency $\omega_3$ can coherently transfer its
excitation to a pair of qubits (1 and 2) if $\omega_1 + \omega_2 =
\omega_3$. The results presented here open the way to {\em
nonlinear optical processes without real photons}. Instead,
virtual photons, which are {\em not} subject to cavity-induced
losses and decoherence, drive the interaction between spatially
separated and nondegenerate qubits, while the qubit-qubit resonant
excitation transfer can be well  described by the Tavis-Cummings
(TC) model \cite{Tavis1968}, the process proposed here cannot be
described without including the counter-rotating terms in the
atom-field interaction Hamiltonian, neglected in the TC model.
Although the Hamiltonian of a realistic atom-cavity system
contains counter-rotating terms (allowing the simultaneous
creation or annihilation of an excitation in both atom and cavity
mode), these terms can be safely neglected for coupling rates that
all small with respect to the atomic transition frequency and the
cavity-mode resonance frequency. However, when the coupling rate
increases, the counter-rotating terms start to play an important
role, giving rise to a new regime of cavity quantum
electrodynamics (QED). This ultrastrong-coupling (USC) regime was
recently realized in a variety of solid-state
systems~\cite{Gunter2009,
FornDiaz2010,Todorov2010,FornDiaz2016,Niemczyk2010,
Schwartz2011,Scalari2012,Geiser2012,Kena-Cohen2013,Gambino2014,Maissen2014,Goryachev2014,
Baust2016,yoshihara2016,Langford2016,Chen2017,Braumuller2016}.
This opens the door to the study of the physics of virtual
processes that do not conserve the number of
excitations~\cite{DeLiberato2009,
Ai2010,Stassi2013,Ridolfo2013,Garziano2013,Garziano2014,Huang2014,Zhao2015,Kockumb2017}.
Recently, it has been shown that, through higher-order processes,
where virtual photons are created and annihilated, an effective
deterministic coupling between two states of such a system can be
created giving rise to new effects such as multiphoton Rabi
oscillations~\cite{Zhu2013,Law2015,Garziano2015} and a single
photon exciting multiple atoms~\cite{Garziano2016}. Moreover, it
has been shown that almost any analog of nonlinear optical
processes is feasible~\cite{Kockum2017} without the need for
high-intensity driving fields. The results presented here go one
remarkable step forward, beyond Ref.~\cite{Kockum2017}, showing
that nonlinear optical processes involving only atoms, without the
need for {\em real} photons, are also feasible.

In quantum nonlinear optics, the effective interaction Hamiltonian
for a three-mode parametric process can be written
as~\cite{mandel1995optical}
\begin{equation} \label{Hno}
\hat V_{\rm c}^{(3)} = K^{(3)}\, \hat a^\dag_1 \hat a^\dag_2 \hat
a_3 + {\rm H.c.}\, ,
\end{equation}
where $\hat a_i$ and $\hat a^\dag_i$ are photon destruction and
creation operators for the $i$th mode, and $K^{(3)}$ is a constant
describing the strength of the nonlinear interaction. This
Hamiltonian describes well-known nonlinear optical processes, such
as sum-frequency generation and spontaneous parametric
down-conversion. The latter process is used especially as a source
of entangled photon pairs and of single photons. Analogously, the
process  proposed  here can be described by the effective
three-qubit  Hamiltonian,
\begin{equation} \label{H3}
\hat V^{(3)} = J^{(3)}\, \hat \sigma_+^{(1)} \hat \sigma_+^{(2)}
\hat \sigma_-^{(3)} + {\rm H.c.}\, ,
\end{equation}
where $\hat \sigma^{(i)}_\pm$ are the raising (+) and lowering (-)
Pauli operators for qubit $i$, and the effective coupling
$J^{(3)}$, as we show, can be calculated by perturbation theory.
This three-body effective interatomic interaction describes
three-qubit mixing (3QM) processes as  the coherent and reversible
transfer of an excitation from one qubit to two
spatially-separated qubits or vice versa. We show that four-qubit
mixing (4QM), described by the following effective Hamiltonian
\begin{equation} \label{H4}
\hat V_I^{(4)} = J^{(4)}\, \hat \sigma_-^{(1)} \hat \sigma_-^{(2)}
\hat \sigma_+^{(3)} \hat \sigma_+^{(4)} + {\rm H.c.}\, ,
\end{equation}
is also possible, where $J^{(4)}$ is the effective coupling
strength. The four-wave mixing of light~\cite{shen1984principles}
arises from  third-order optical nonlinearities. In typical cases,
a photon of frequency $\omega_4$ results from the annihilation of
photons at $\omega_1$ and $\omega_2$ and the stimulated emission
of one at $\omega_3$ with $\omega_4 + \omega_3 = \omega_1 +
\omega_2$. The process can also be spontaneous, occurring even in
the absence of stimulation at $\omega_3$. The effective potential
in Eq.~(\ref{H4}) enables the simultaneous excitation transfer
from qubits 1 and 2 to qubits 3 and 4 when the qubit transition
frequencies satisfy the relation $\omega_1 + \omega_2 = \omega_3+
\omega_4$, which is an interatomic scattering  process without the
presence of real photons. This process is the qubit analog of the
spontaneous four-wave mixing of light (see, e.g.,
Ref.~\cite{takesue2004}). In the following we refer to it as
type-I 4QM, in order to distinguish it from a different type of
4QM (type II), achievable in our system in the USC regime when
$\omega_4 = \omega_1 + \omega_2 + \omega_3$. This latter process
can efficiently perform quantum repetition codes after a proper
evolution time. We show that such spontaneous evolution of the
system can be used for error-correction
codes~\cite{Shor95,PerryBook} for encoding, decoding, and
error-syndrome detection.

Note that the framework proposed here is different from nonlinear
{\em atom optics}~\cite{Rolston2002}, where the atomic
center-of-mass degree of freedom is involved. Coherent matter
waves in the form of Bose-Einstein condensates have led to the
development of nonlinear and quantum atom optics, where atomic
waves are manipulated in a manner analogous to the manipulation of
light~\cite{Lenz1993,Trippenbach1998}. For example, coherent
four-wave mixing (in which three sodium matter waves of differing
momenta mix to produce a fourth wave with another momentum) has
been demonstrated experimentally~\cite{Deng1999}.

It has also been shown that a system of trapped ions can be used
to implement spin models with three-body
interactions~\cite{Bermudez2009}. However, in contrast to the
framework proposed here, where the effective interaction is
enforced by the field quantum vacuum only, in trapped ions the
effective  three-spin interactions are induced by external lasers.

As we show below, the effective Hamiltonians in Eqs.~(\ref{H3})
and (\ref{H4}) can be derived  from a generalized Dicke
Hamiltonian~\cite{Dicke1954}, describing three or more qubits
interacting with the same oscillator (cavity mode). The
interaction Hamiltonian also includes a longitudinal coupling
term~\cite{Liu2005}, which arises when the inversion symmetry of
the potential energy of the artificial atoms (qubits) is
broken~\cite{You2011,Gu2017}.

\section{Results}
\label{results}
\subsection{Description of the system}
Here we examine a quantum system of $N$ two-level atoms (with
possible symmetry-broken potentials) coupled to a single-mode
resonator.
The Hamiltonian describing this system  is (e.g.,
Refs.~\cite{Niemczyk2010,Garziano2014}):
\begin{equation}
\hat H_0 =  \hat H_{\rm q} + \hat H_{\rm c} +  \hat V\, ,
\label{H0}
\end{equation}
where (using $\hbar =1$) $\hat H_{\rm q} =  \sum_i (\omega_i/2)\,
\hat \sigma^{(i)}_z $ and $ \hat H_{\rm c} =\omega_{\rm c}\, \hat
a^\dag \hat a$ describe the qubit and cavity Hamiltonians,
respectively, in the absence of interaction. The qubits-cavity
interaction is \be \hat V= \hat X \sum_i \lambda_i (\cos
\theta_i\,  \hat \sigma^{(i)}_x + \sin \theta_i\, \hat
\sigma^{(i)}_z )\, , \label{eq:V3QM} \ee where $\hat X =  \hat a +
\hat a^\dag$, $\hat \sigma^{(i)}_x$ and $\hat \sigma^{(i)}_z$ are
Pauli operators for the $i$th qubit, $\lambda_i$ are the coupling
rates of each qubit to the cavity mode, and $\theta_i$ are
parameters determining the relative contribution of the transverse
and longitudinal couplings. For $\theta_i = 0$, the parity of
qubit $i$ is conserved. For flux qubits, the angles $\theta_i$, as
well as the transition frequencies $\omega_i$, can be continuously
and individually tuned by changing the external flux biases
\cite{Liu2005,Niemczyk2010}. In contrast to the TC model, the
Hamiltonian in Eq.~(\ref{H0}) explicitly contains counter-rotating
terms of the form $\hat \sigma^{(i)}_+ \hat a^\dag$, $\hat
\sigma^{(i)}_- \hat a$, $\hat \sigma^{(i)}_z \hat a^\dag$, and
$\hat \sigma^{(i)}_z \hat a$. The first (second) term creates
(destroys) two excitations, while the third (fourth) term creates
(destroys) one excitation. Equation~(\ref{H0}) represents the
simplest Hamiltonian describing the interaction of $N$ atoms (with
possible symmetry-broken potentials) with the electromagnetic
field of a cavity beyond the rotating-wave approximation (RWA).
This model is well suited for describing atoms with very large
anharmonicity, such as flux qubits. However, we expect that the
results presented below also apply, at least qualitatively, to
more general atom-cavity systems, when additional atomic levels
are involved.

In the case of two qubits ($i=1, 2$), dropping the
counter-rotating terms, considering the dispersive regime
($|\Delta_i| = |\omega_{i}- \omega_{\rm c}| \gg \lambda_i$), and
applying second-order perturbation theory, it is possible to
derive from the atom-cavity interaction Hamiltonian [see
Eq.~\ref{H0}], the following effective interaction Hamiltonian
\cite{Majer2007, Blais2007}
\begin{equation}
\label{Heff2} \hat V^{(2)} = J^{(2)}\, \hat \sigma^{(2)}_+ \hat
\sigma^{(1)}_- + {\rm H.c.}\, ,
\end{equation}
where $J^{(2)} =\lambda_1\, \lambda_2 (1/ \Delta_1 + 1/
\Delta_2)/2$. In this regime, no energy is exchanged between the
qubits and the cavity. This qubit-qubit interaction is the result
of virtual exchange of photons with the cavity. It gives rise to a
qubit-qubit avoided-level crossing when the transition energy of
one of the two qubits is continuously tuned across the fixed
transition energy of the other. When the qubits are degenerate, an
excitation in one qubit can be transferred to the other qubit by
virtually becoming a photon in the cavity. However, when the
qubits are nondegenerate, $| \omega_1 - \omega_2 | \gg J^{(2)}$,
this process does not conserve energy and, therefore, the
interaction is effectively turned off.

The dispersive-regime condition $|\Delta_i| = |\omega_{i}-
\omega_{\rm c}| \gg \lambda_i$ is necessary to enable the virtual
exchange of photons. Moreover, in order to ensure that only a
negligible population of real photons is present, the atom-cavity
detuning has to be much larger than the atomic and photonic
decoherence rates.

Throughout this paper, we consider a single-mode optical
resonator. Many circuit-QED experiments use an $LC$ resonator,
which only has a single mode.  When additional modes are
considered, the dispersive-regime conditions have to hold for all
the modes. Defining $|\Delta_{i,n}| =  |\omega_{i} - \omega_{{\rm
c},n}|$, where $\omega_{{\rm c},n}$ is the $n$th mode frequency,
the conditions are $|\Delta_{i,n}| \gg \lambda_i$ and
$|\Delta_{i,n}| \gg \kappa_n$, where  $\kappa_n$ is the decay rate
of the $n$th mode. If these conditions are satisfied, following
the procedure of Ref.~\cite{Blais2007}, it is straightforward to
find $J^{(2)} = \sum_n [\lambda_{1,n}\, \lambda_{2,n} (1/
\Delta_{1,n} + 1/ \Delta_{2,n})/2]$. This series is expected to
converge, owing to the suppression of  light-matter coupling
$\lambda_{i,n}$ at high frequencies (see, e.g.,
Ref.~\cite{malekakhlagh2017}). If the lowest-frequency mode is not
too far detuned from the atomic transition frequencies, and the
modes are well separated spectrally,  only the lowest-frequency
mode provide a significant contribution. The experimental results
shown in Ref.~\cite{Majer2007}, obtained using a $\lambda/4$
coplanar waveguide-resonator, are very well described considering
a single mode, since higher energy  modes are too far off
resonance to give a significant contribution. Analogous
considerations apply to the processes described here. On the
contrary, in  a large cavity ($l \gg \lambda_i$)  or in a
transmission line, the modes constitute either a quasicontinuum or
a continuum, and the single-mode description adopted here does not
work. In this case, one possibility to realize a qubit-qubit
interaction mediated by virtual photons  is to consider atoms with
transition frequencies outside the frequency bandwidth of the
photonic system.

In the following, we show how the  Hamiltonian in Eq.~(\ref{H0})
can also give rise to effective qubit-qubit interactions involving
more than two qubits. Specifically, we consider nonlinear
interatomic processes involving nondegenerate qubits, such as 3QM
and 4QM.
\subsection{Three-qubit mixing}
\begin{figure}
    \centering
    \includegraphics[width = \linewidth]{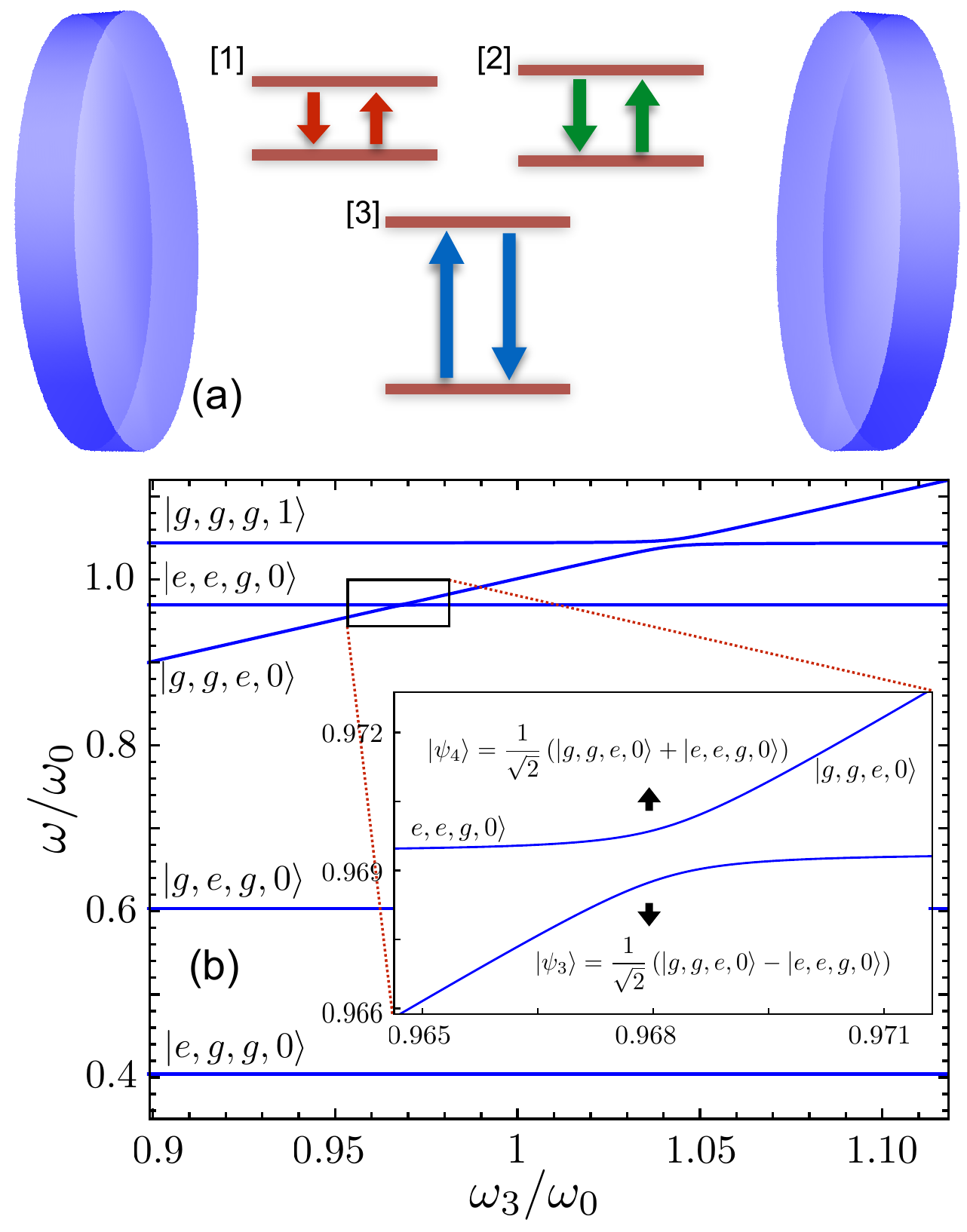}
    \caption{(a) Schematic representation of three nondegenerate qubits interacting with the electromagnetic field of a cavity.
    (b) Lowest-energy levels, indicated with $\omega$ ($\hbar =1$), of the system constituted by three qubits interacting with a cavity mode as a function of the normalized frequency of qubit 3, obtained by numerically diagonalizing the Hamiltonian in Eq.~(\ref{H0}). The transition frequencies of the other two qubits, as well as the resonance frequency of the cavity mode, are kept fixed. All the  parameters used here are specified in the text. The enlarged view of the boxed {\em apparent} crossing in the inset displays a clear anticrossing level splitting. When  the splitting is at its minimum, the eigenstates of the system are approximately  symmetric and antisymmetric superpositions of the states $|g,g,e,0 \rangle$ and $|e,e,g,0 \rangle$.
        \label{fig:1}}
\end{figure}
Here we consider three nondegenerate qubits of transition
frequencies $\omega_1 \neq \omega_2 \neq \omega_3$ coupled to a
cavity mode [see \figref{fig:1}(a)]. Figure\,\ref{fig:1}(b) shows
the energy levels for the excited lowest-energy states as a
function of the frequency of qubit 3, obtained by numerically
diagonalizing Eq.~(\ref{H0}). For each value of $\omega_3$, the
energy scale is chosen such that the ground-state energy is equal
to zero. All the parameters are provided in terms of a fixed
reference frequency $\omega_0$. We assume that qubits 1 and 2 are
ultrastrongly coupled to the cavity mode, while the coupling
strength of qubit 3 with the cavity mode is lower. We used
$\lambda_3/\omega_0 = 5 \cdot 10^{-3}$,  $\lambda_1/\omega_0 =
\lambda_2/\omega_0 = 0.13$, $\theta_i = \pi/6$ for all the qubits,
$\omega_1/ \omega_0 = 0.4$, $\omega_2 /\omega_0 = 0.6$, and
$\omega_{\rm c}  = \omega_0  + 2.5 \lambda_1$. The lowest-energy
horizontal line (with $\omega/ \omega_0 \approx 0.4$) corresponds
to the excitation of qubit 1. We indicate this state as $|
\psi_1\rangle = |e,g,g,0 \rangle$, where the first three entries
in the ket describe the states of the three qubits and the last
entry describes the cavity-mode state. We observe that this
eigenstate, corresponding to the excitation of the {\em physical}
qubit 1, can differ from the bare state $|e,g,g,0 \rangle_{\rm
b}$, describing the excitation of qubit 1 in the absence of its
interaction with the cavity mode (see, e.g.,
Ref.~\cite{DiStefano2016}). Owing to the dressing effects induced
by the counter-rotating terms in the interaction Hamiltonian,
differences between bare and physical states occur for all the
energy eigenstates.   The second  horizontal line (with $\omega/
\omega_0 \approx 0.6$) corresponds to $| \psi_2\rangle = |g,e,g,0
\rangle$. A signature of the discussed dressing is the slight
frequency shift occurring between the bare qubit frequencies
$\omega_2$ and $\omega_3$ and the two lowest-energy (horizontal)
levels displayed in \figref{fig:1}(b) [with $\omega/ \omega_0
\approx 0.4$ and $0.6$] corresponding to the {\em physical}
transition frequencies of qubit 1 and 2 (in the presence of the
interaction with the cavity mode), respectively. In the region
much below the first (apparent) crossing ($\omega_3 / \omega_0 <
0.95$), the third level corresponds to $| \psi_3\rangle = |g,g,e,0
\rangle$, as can  also be inferred from its linear dependence on
$\omega_3$. In the same frequency region, the fourth level (with
$\omega/ \omega_0 \approx 0.97$) corresponds to the simultaneous
excitation of qubits 1 and 2, $| \psi_4 \rangle = |e,e,g,0
\rangle$, while the fifth level (with $\omega/ \omega_0 \approx
1.04$) corresponds to the one-photon state $| \psi_5 \rangle =
|g,g,g,1 \rangle$. When increasing $\omega_3 /\omega_0$, the
energy level associated with $|g,g,e,0 \rangle$ rises, and it
reaches the energy levels corresponding to $|e,e,g,0 \rangle$ and
$|g,g,g,1 \rangle$. When the (dressed) energy of qubit 3
approaches that of the cavity mode, a clear anticrossing can be
observed. This is the ordinary vacuum Rabi splitting, which can
also be  reproduced within the RWA. When this splitting is at its
minimum, the eigenstates of the system are the symmetric and
antisymmetric superpositions of the states $|g,g,g,1 \rangle$ and
$|g,g,e,0 \rangle$, as confirmed by numerical calculations. More
interestingly, \figref{fig:1}(b) also displays an apparent
crossing at lower energy when $\omega_3 \simeq \omega_1 +
\omega_2$ ($\approx 0.4+0.6$). The enlarged view in the inset
shows that this is actually an anticrossing level splitting. When
this splitting is at its minimum (see inset in
Fig.~\ref{fig:1}(b), the two system eigenstates $\ket{\psi_{3,4}}$
are, respectively, the antisymmetric and symmetric superpositions
of the states $\ket{g,g,e,0}$ and $\ket{e,e,g,0}$. This avoided
level crossing indicates a coherent coupling, which does not
conserve the number of excitations, of the  three
spatially-separated qubits.


The origin of this coupling can be understood using time-dependent
fourth-order perturbation theory, identifying the resulting
transition amplitude between the initial state $| i \rangle \equiv
|e,g,g,0 \rangle_{\rm b}$  and the final state $| f \rangle \equiv
|g,e,e,0 \rangle_{\rm b}$ (or vice versa) with the effective
coupling strength determining this level splitting. According to
fourth-order perturbation theory, this coupling can be expressed
as~\cite{Kockum2017}
\begin{equation}\label{J3}
\lambda_{\rm eff}  =
 \sum_{n,m,k} \frac{V_{fn} V_{nm} V_{mk} V_{ki}}{\left( E_i - E_n \right) \left( E_i - E_m \right) \left( E_i - E_k \right)}\, ,
\end{equation}
where  $V_{n,m} = \langle n | \hat V | m \rangle$. Although the
initial and final states do not contain photons,  the coupling is
determined by the interaction of the qubits with the cavity field.
States with nonzero photon number play a role only as intermediate
levels ($|n \rangle$, $|m \rangle$, and $|k \rangle$) reached by
virtual transitions (see diagrams in \appref{app:3Qubits}). The
two states $| i \rangle$  and  $| f \rangle$ are connected via a
fourth-order process and there are no lower-order contributions.
There are 48 paths connecting the states to this order, as shown
in \figref{fig:48Paths} in \appref{app:3Qubits}. These paths
clearly show that the three qubits are connected by a nonlinear
optical process involving {\em only} virtual photons. This
analysis has been performed for two-level atoms, which is a good
approximation for flux qubits where the next higher level can be
energetically very far (see, e.g., Ref.~\cite{Chiorescu2004}). Of
course, additional paths must  be taken into account for
multilevel systems. The resulting effective coupling between the
states $|g,g,e,0 \rangle$ and $|e,e,g,0 \rangle$ can be described
by the effective Hamiltonian in Eq.~(\ref{H3}), where $J^{(3)} =
\lambda_{\rm eff}$ [corresponding to half the minimum level
splitting shown in the inset in \figref{fig:1}(b)].

\begin{figure}[ht!]
    \centering
    \includegraphics[scale=0.8]{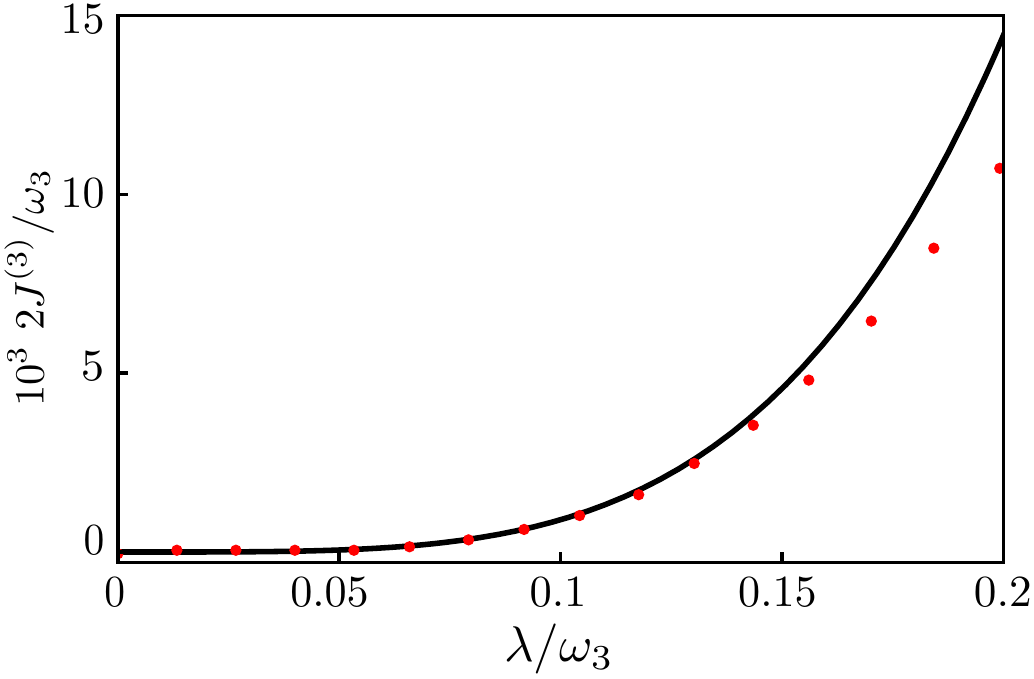}
    \caption{Comparison between the numerically-calculated normalized effective Rabi splitting (points)  corresponding to $J^{(3)}$ in \appref{app:3Qubits} and the corresponding calculations using the fourth-order perturbation theory (continuous black curve), as a function of the normalized interaction strength $\lambda/\omega_3$.
        \label{fig:2}}
\end{figure}

The analytical expression obtained from Eq.~(\ref{J3}), which
considers the 48 contributions, is very cumbersome. However, it
can be simplified considerably if we assume $\lambda_1 = \lambda_2
= \lambda_3 = \lambda$ and $\omega_1 = \omega_2 = \omega_3/2$. In
this case, the final result is \be J^{(3)} = \frac{64 \lambda^4
\omega_c^2 \left(4\omega_c^2 - 7\omega_3^2 \right) \sin \theta
\cos^3 \theta}{\omega_3\left(\omega_3^2 - \omega_c^2
\right)\left(\omega_3^2 - 4\omega_c^2 \right)^2}.
\label{eq:GeffSimpCD} \ee We note that the effective coupling
becomes zero when $\omega_c = \frac{\sqrt{7}}{2} \omega_3$. It is
not easy to see how the interference between the 48 paths becomes
destructive for this particular value of $\omega_c$. Looking at
the denominator of \eqref{eq:GeffSimpCD}, we see that $J^{(3)} \to
\infty$ when $\omega_c \to \omega_3$ or $\omega_c \to \frac{1}{2}
\omega_3$, i.e., when the cavity becomes resonant with one of the
qubits. Perturbation theory is not valid around those points. We
also note that the coupling, also in the unsimplified general
case, is proportional to $\sin\theta \cos^3\theta$, which implies
that the maximum coupling is achieved when $\theta = \pi/6$.
Figure~\ref{fig:2} displays a comparison of the magnitudes of the
effective Rabi splitting $2  J^{(3)}/ \omega_{3}$  obtained
analytically [Eq.~(\ref{eq:GeffSimpCD})] via fourth-order
perturbation theory (black continuous curve) and by numerical
diagonalization of the Hamiltonian in Eq.~(\ref{H0}), as a
function of  the normalized interaction strength $\lambda /
\omega_{3}$ ($\lambda = \lambda_{1,2,3}$). The other parameters
are the same as those used to obtain the results in
Fig.~\ref{fig:1}. The agreement is very good for normalized
interaction strengths $\lambda/ \omega_3 \lesssim 0.15$.
Figure~\ref{fig:2} shows that an effective coupling rate  well
beyond  typical decoherence rates of circuit-QED systems (see,
e.g., Ref.~\cite{Yan2016}) can be obtained  already at a coupling
strength $\lambda/\omega_3 \sim 0.1$.

In order to demonstrate the simultaneous excitation transfer from
qubit 3 to qubits 1 and 2, we fix $\omega_3 \simeq \omega_1 +
\omega_2$, so that the system is at its minimum-level splitting
[see inset in \figref{fig:1}(b)]. The minimum occurs when the
transition energy of the dressed qubit 3 is equal to the energy
level corresponding to the simultaneous excitation of the dressed
qubits 1 and 2. We  study the dynamics after initial preparation
of the system in the symmetric superposition of the eigenstates
associated with the two split energy levels: $(| \psi_3 \rangle +
| \psi_4 \rangle)/ \sqrt{2} = |g,g,e,0 \rangle$, corresponding to
the excitation of the dressed qubit 3. The system can be prepared
in this state by directly exciting the detuned qubit 3 by sending
a $\pi$ electromagnetic pulse, followed by a  flux shift  that
brings the qubit into resonance (minimum splitting). Another
possibility is to start with qubit 3 already on resonance and
excite it by  a  $\pi$ pulse much faster than $\pi / J^{(3)}$.

\begin{figure}[ht!]
    \centering
    \includegraphics[scale=0.73]{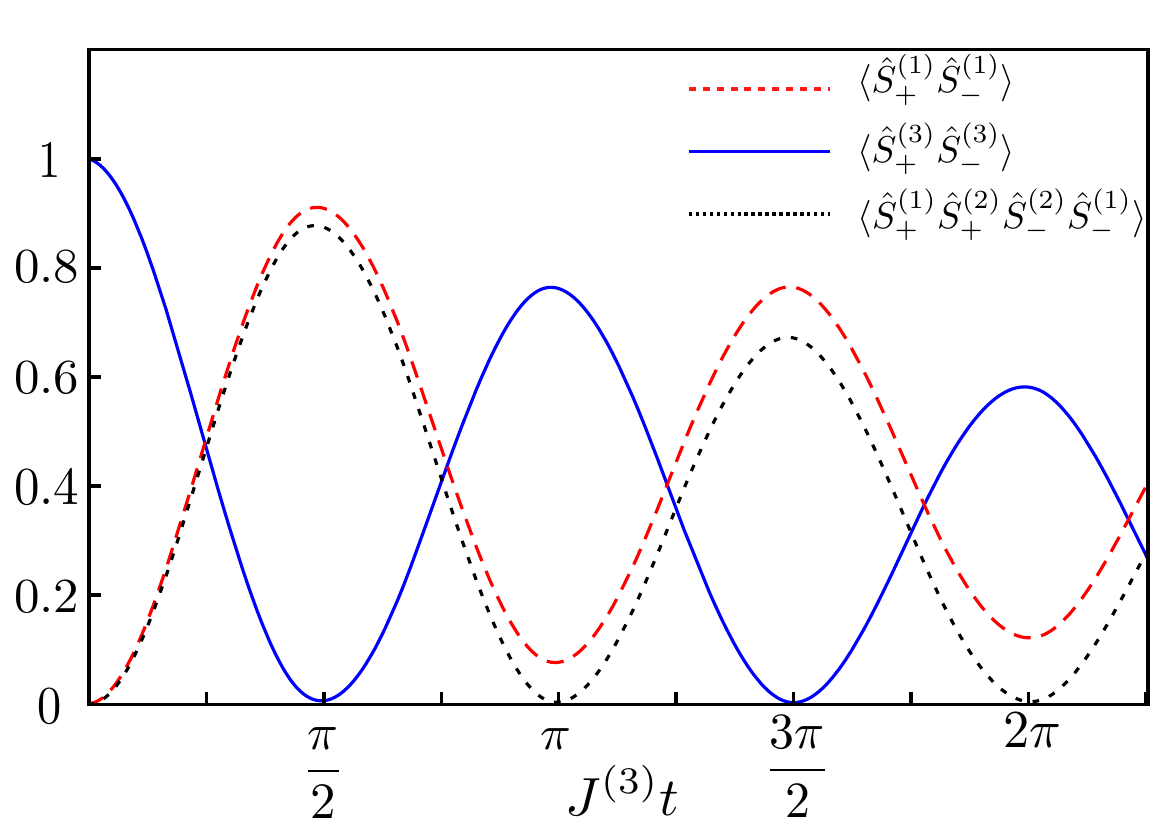}
    \caption{Time evolution of the mean excitation probabilities of qubits 1 and 3, $\langle \hat S_+^{(1)} \hat S^{(1)}_- \rangle$ and $\langle \hat S_+^{(3)} \hat S^{(3)}_- \rangle$, and of the two-qubit correlation function $\langle \hat S_+^{(1)} \hat S^{(2)}_+ \hat S^{(2)}_-\hat S^{(1)}_- \rangle$. The initial state is $|g,g,e,0 \rangle$.
        \label{fig:3}}
\end{figure}

The influence of the cavity-field damping and atomic decay on the
process can be studied by the master-equation approach in the
dressed picture~\cite{Beaudoin2011, Ridolfo2012}. We consider the
system interacting with zero-temperature baths. The master
equation is obtained by using the Born-Markov approximation
without the post-trace RWA~\cite{Law2015}. We study the dynamics
of the relevant system population and correlation functions by
introducing the dressed-qubit lowering and raising operators
$\hat S^{(i)}_-$ and $\hat S^{(i)}_+ = (\hat S^{(i)}_-)^\dag$.
They are a direct generalization of $\hat \sigma^{(i)}_-$ and
$\hat \sigma^{(i)}_+$. For qubit 1: \be \hat S^{(1)}_-
=\sum^\infty_{n= 0}  \sum^e_{\alpha,\beta= g}
|g,\alpha,\beta,n\rangle \langle e,\alpha,\beta,n|\, . \ee In the
absence of interaction, when  $|p,q,r,n\rangle$ are bare states,
the operators $\hat S_-^{(i)}$ coincide with the usual lowering
Pauli operators $\hat \sigma_-^{(i)}$. Figure\,\ref{fig:3}
displays the time evolution of the mean excitation probability
 $\langle \hat S_+^{(i)} \hat S^{(i)}_- \rangle$
 of qubits 1 and 3 (the dynamics of qubit 1 coincides with that of qubit 2), as well as that of the two-qubit correlation function
$\langle \hat S_+^{(1)} \hat S^{(2)}_+ \hat S^{(2)}_-\hat
S^{(1)}_- \rangle$. The initial state is $|g,g, e,0 \rangle =
(|\psi_3 \rangle + \psi_4 \rangle)/\sqrt{2}$. The parameters are
those used to obtain the energy levels in \figref{fig:1}(b). For
the decay rates of the qubit ($\gamma$) and the cavity ($\kappa$),
we used $\gamma = \kappa = 3 \times 10^{-5} \omega_0$ . Quantum
Rabi oscillations, showing the reversible excitation exchange
between qubit 3 and qubits 1 and 2, can be clearly observed. Note
that, during the time evolution displayed in Fig.~\ref{fig:3}, the
photon population (not shown) reaches a maximum value of $ \sim
1.5 \times 10^{-2}$. This small population decreases rapidly  when
increasing the detuning $\Delta_3$ between the cavity-mode and
qubit 3. We also checked that increasing the photon damping,
provided that $\kappa \lesssim  2 \Delta_3$, does not affect the
displayed dynamics. After half a Rabi period ($ t =
\pi/2J^{(3)}$), the excitation is fully transferred from qubit 3
to qubits 1 and 2, which reach their maximum excitation
probability. The presence of damping prevents $\langle \hat
S_+^{(1)} \hat S^{(1)}_- \rangle$ from reaching $1$. We observe
that the single-qubit excitations and the two-qubit correlation
function
$\langle \hat S_+^{(1)} \hat S^{(2)}_+ \hat S^{(2)}_-\hat
S^{(1)}_- \rangle$
  almost coincide at early times. This almost-perfect two-qubit correlation is a clear signature of the joint excitation of qubits 1 and 2: if one qubit becomes excited, the probability that  the other one is excited is also very close to one. However, as expected, the two-qubit correlation is more fragile to losses.
We also observe that, at time $t =  \pi/ 4 J^{(3)}$, this process
spontaneously gives rise to the maximally entangled three-qubit
state $(|g,g,e \rangle -i |e,e,g \rangle)/\sqrt{2}$ when damping
can be neglected (the factorized photonic vacuum state has been
disregarded). This state is the Greenberger-Horne-Zeilinger (GHZ)
state, up to a local transformation.

We observe that, during the time evolution displayed in
Fig.~\ref{fig:3}, the photon population (not shown) reaches a
maximum value of $ \sim 1.5 \times 10^{-2}$. This small population
decreases rapidly  when increasing the detuning $\Delta_3$ between
the cavity mode and qubit 3. We also checked that increasing the
photon damping, provided that $\kappa \lesssim  2 \Delta_3$, does
not affect the displayed dynamics. This result shows that the 3QM
process is not influenced by the cavity-loss rate $\kappa$, owing
to the negligible probability to have real photons in the cavity.
This result, however, does not imply a total immunity to losses of
the quantum bus. For example, if an impurity atom, almost resonant
with the transition energy of qubit 3, is present inside the bus,
the excitation would be partly transferred to the impurity atom.

This 3QM offers also the possibility to encode an arbitrary qubit
state $a |g \rangle + b |e \rangle$ into a two-qubit entangled
state so that $(a |g \rangle + b |e \rangle)|g \rangle | g \rangle
\to |g \rangle (a |g \rangle |g \rangle -i b |e \rangle|e
\rangle)$. This operation can be realized by just letting the
system evolve spontaneously for a time $t = \pi/ 2 J^{(3)}$.

\subsection{Four-qubit mixing}
\begin{figure*}
    \centering
    \includegraphics[scale=1.]{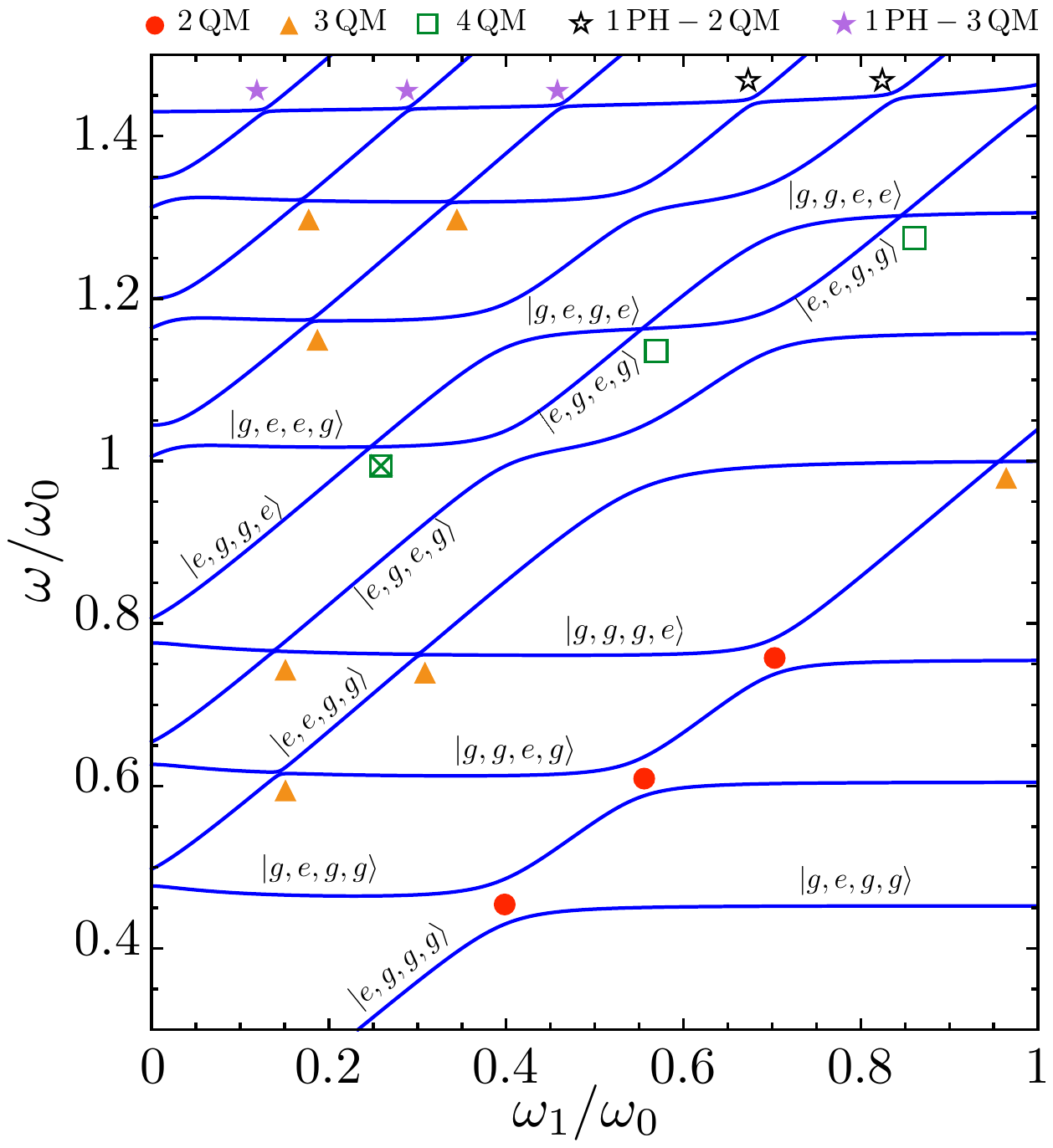}
    \caption{Four nondegenerate qubits dispersively coupled to a cavity mode at higher energy.
        Lowest energy levels of the system  as a function of the frequency of qubit 1, obtained by numerical diagonalization of the Hamiltonian in Eq.~(\ref{H0}) with four qubits. The transition frequencies of the other three qubits as well as the resonance frequency of the cavity mode are kept fixed. All the parameters used are provided in the text.
        \label{fig:4}}
\end{figure*}

Here we consider four nondegenerate qubits coupled to a cavity
mode in the dispersive regime and investigate the 4QM process.
Figure~\ref{fig:4} shows the energy levels for the lowest-energy
excited states as a function of the frequency of qubit 1, obtained
by numerically diagonalizing the Hamiltonian in Eq.~(\ref{H0}).
Also in this case, for each value of $\omega_1$, the energy scale
is chosen such that the ground-state energy is equal to zero. All
the values are provided in terms of a fixed reference frequency
$\omega_0$. We used $\lambda_i/\omega_0  = 0.15$ and $\theta_i =
\pi /6$ for all the qubits. We also set the transition frequencies
of qubits 2, 3, and 4 as  $\omega_2/ \omega_0 = 0.4$, $\omega_3
/\omega_0 = 0.55$, $\omega_4 /\omega_0 = 0.7$, and the resonance
frequency of the cavity mode as $\omega_{\rm c}/\omega_0 =1.4$.

\begin{figure}
    \centering
    \includegraphics[width=\linewidth]{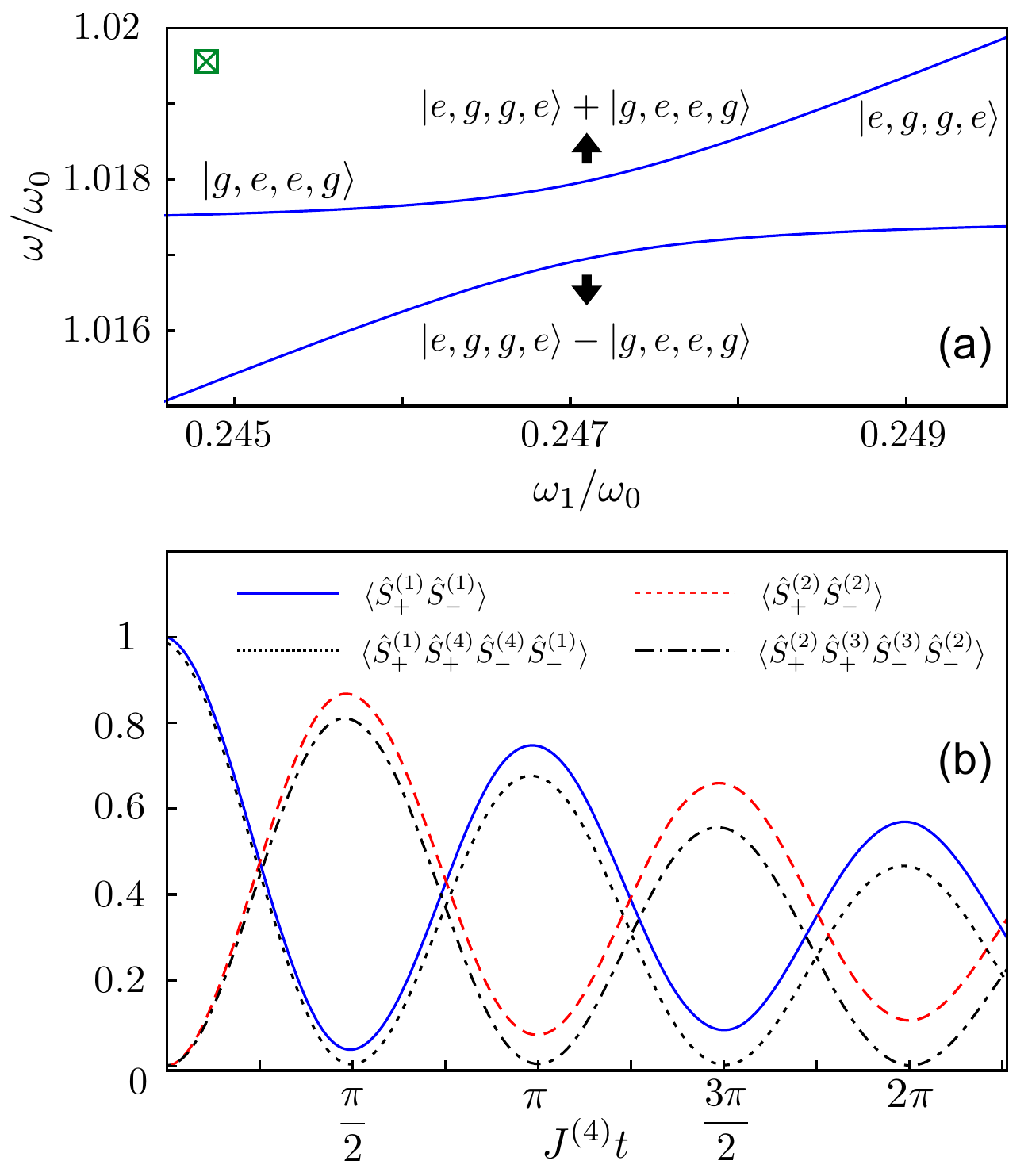}
    \caption{(a) Enlarged view of the {\em apparent} crossing {\color{ao(english)}$\boxtimes$} in \figref{fig:4} determining a type-I 4QM. (b) Type-I 4QM: Time evolution of the mean excitation probability of qubits 1 and 2,
     $\langle \hat S_+^{(1)} \hat S^{(1)}_- \rangle$ and $\langle \hat S_+^{(2)} \hat S^{(2)}_- \rangle$
     , and of the two-qubit correlation functions
 $\langle \hat S_+^{(1)} \hat S^{(4)}_+ \hat S^{(4)}_-\hat S^{(1)}_- \rangle$ and $\langle \hat S_+^{(2)} \hat S^{(3)}_+ \hat S^{(3)}_-\hat S^{(2)}_- \rangle$. The initial state is $|e,g,g,e,0 \rangle$.
        \label{fig:5}}
\end{figure}

The figure displays several {\em apparent} level crossings and
anticrossings, corresponding to different kinds of normal-mode
couplings. The avoided level crossings indicated by the red
circles {\color{red} {\Large $\bullet$}} originate from two-qubit
resonant interactions described by the effective Hamiltonian in
Eq.~(\ref{Heff2}). In the time domain, this type of interaction
leads to two-qubit mixing (2QM). The {\em apparent} level
crossings labeled by the orange triangles
{\color{orange}$\blacktriangle$} give rise to 3QM. For example,
the lowest-energy orange triangle (at $\omega/\omega_0 \simeq
0.6$) marks the coupling between the states $|g,g,e,g,0\rangle$
and $|e,e,g,g,0 \rangle$. The stars in the higher-energy region
describe the coupling of a single photon to two ({\large
\text{\ding{73}}}) or three  ({\color{amethyst}$ \bigstar$})
qubits, studied in Ref.~\cite{Garziano2016}. Type-I 4QM processes
are labeled by {\color{ao(english)}$\square$}. An enlarged view of
the {\em apparent} crossing {\color{ao(english)}$\boxtimes$} in
\figref{fig:4}  is plotted in  \figref{fig:5}(a). There, an
anticrossing level splitting $\sim 10^{-3} \omega_0$ can be
observed. When  the splitting is at its minimum, the two  system
eigenstates are the symmetric and antisymmetric superpositions of
the states $|e,g,g,e,0 \rangle$ and $|g,e,e,g,0 \rangle$. This
avoided-level crossing is the signature of a four-qubit coherent
coupling.

Also in this case, the origin of this coupling can be understood
using time-dependent fourth-order perturbation theory, using
Eq.~(\ref{J3}) and considering the states $|e,g,g,e,0 \rangle_{\rm
b}$ and $|g,e,e,g,0 \rangle_{\rm b}$ as the initial and final
states (see \appref{analqubits}). We notice that type-I 4QM, in
contrast to the 3QM process, conserves the number of excitations.
Hence we can expect that it can be described within the RWA (for
the TC model). However, by numerically diagonalizing the TC model
we find no avoided level splitting. As shown in
\appref{analqubits}, the fourth-order perturbation theory shows
that the coupling goes to zero owing to perfect cancellation
between the different contributions to the matrix element. This
4QM process can be described by the effective interaction
Hamiltonian in Eq.~(\ref{H4}), which determines  the coupling
between the states  $|g,g,e,e,0 \rangle$ and $|e,e,g,g,0 \rangle$.

Figure \ref{fig:5}(b) displays the time evolution of the mean
excitation probability of qubits 1 and 2 (the dynamics of qubits 3
and 4 coincide with that of qubit 1 and 2, respectively), as well
as that of the two-qubit correlation functions
$\langle \hat S_+^{(1)} \hat S^{(4)}_+ \hat S^{(4)}_-\hat
S^{(1)}_- \rangle$ and $\langle \hat S_+^{(2)} \hat S^{(3)}_+ \hat
S^{(3)}_-\hat S^{(2)}_- \rangle$. The parameters are those used to
obtain the energy levels in \figref{fig:4}. We also set $\gamma =
\kappa = 3 \times 10^{-5} \omega_0$. The initial state is
$|e,g,g,e,0 \rangle$. The system can be prepared in this state by
setting the qubit frequencies such that $\omega_1 + \omega_4 \neq
\omega_2 + \omega_3$ and directly exciting qubits 1 and 4 by
sending a $\pi$  pulse to each of them,  followed by a flux shift
to one of the four qubits that brings the four-qubit system into
resonance [corresponding to the minimum splitting in
\figref{fig:5}(a)]. At $t=0$, qubits 1 and 4 are excited and the
minimum-splitting condition is satisfied. Figure \ref{fig:5}(b)
clearly demonstrates the excitation transfer $|e,g,g,e,0 \rangle
\to |g,e,e,g,0 \rangle$, which is also reversible as the time
evolution for $ \pi/2 \leq J^{(4)}\, t \leq \pi$ shows. If damping
is absent or negligible, the transfer can be deterministic at
$J^{(4)}\, t = \pi/2$,  and  a maximally-entangled  four-qubit
state (GHZ-type state) is obtained at $J^{(4)}\, t = \pi/4$.

This 4QM process can be used to transfer the entanglement from a
pair of qubits to another spatially-separated pair, initially in a
factorized state. Specifically, if the system is initially
prepared in the two-qubit entangled state $(a |g,g \rangle + b |e,
e \rangle)|g,g \rangle$, after a time $t = \pi/ 2 J^{(4)}$, the
entanglement is transferred from qubits 1 and 2 to qubits 3 and 4:
$(a |g,g \rangle + b |e, e \rangle)|g,g \rangle \to |g, g
\rangle(a |g,g \rangle -i b |e, e \rangle)$.

\begin{figure*}
        \centering
        \includegraphics[width=16cm]{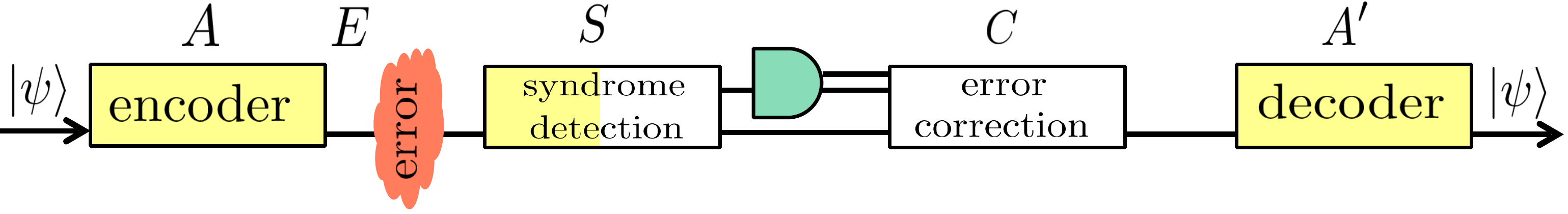}
\caption{Basic modules of a standard error-correction code (ECC).
These include encoding ($A$) and decoding ($A'$) modules (usually
based on quantum repetition codes using multiple CNOT gates), a
dissipative channel or evolution when the error ($E$) happens,
error-syndrome detection ($S$), and error-correction ($C$)
modules. Sometimes, the modules $S$ and $C$ are combined. Here,
single (double) lines denote quantum (classical) channels, and a
half-circle denotes detectors. In \appref{ecc}, we consider in
detail an ECC (see, e.g., Ref.~\cite{PerryBook}) for correcting
either a single bit-flip or phase-flip error. Then, modules $A$
and $A'$ can be entirely implemented (as marked in yellow) by a
type-II 4QM, described by $\hat V_{\rm II}^{(4)}$ for the
evolution time (length) $t'=\pi/(2J^{(4)})$. Typically, modules
$A$ and $A'$ are realized with four CNOT gates in total. Moreover,
the error-syndrome module $S$ can be implemented by 4QM and two
additional CNOT gates, instead of typically four CNOT gates. The
detected error can be corrected by a single-qubit rotation (NOT
gate) applied to a proper qubit, based on the results of the
detectors, as discussed in~\appref{ecc}. }
 \label{ECC1}
    \end{figure*}

Adjusting the transition frequencies of the qubits, a four-qubit
down-conversion (type-II 4QM) analogous to that studied above for
three qubits can also occur. This process is enabled by the
resonant coupling between the states $|e,g,g,g,0 \rangle
\leftrightarrow |g,e,e,e,0 \rangle$ and can be described by the
effective Hamiltonian
\begin{equation} \label{H4p}
\hat V_{\rm II}^{(4)} = J^{(4)}\, ( \hat \sigma_-^{(1)} \hat
\sigma_+^{(2)} \hat \sigma_+^{(3)} \hat \sigma_+^{(4)} + {\rm
H.c.})
\end{equation}
As shown in \appref{S1}, the resulting splitting is of the same
order as that shown in \figref{fig:5}(a). The spontaneous
evolution of the Dicke USC system, effectively described by the
Hamiltonian (\ref{H4p}), corresponding to a type-II 4QM, performs
the transformation
 \beq
(a\ket{g}+b\ket{e})\ket{ggg} \rightarrow
\ket{g}(a\ket{ggg}-ib\ket{eee}),
 \eeq
after the evolution time $t'=\pi/(2J^{(4)})$. This operation
corresponds, up to a single-qubit phase gate,  to a three-qubit
repetition code, which is usually implemented with two
controlled-NOT (CNOT) gates. Repetition codes are basic elements
of error-correction codes (ECCs)~\cite{Shor95,NielsenBook}; in
particular, for encoding $A$ and decoding $A'$, as shown and
explained in Fig.~\ref{ECC1}. Note that the error $E$ can be
corrected in the module $C$ by a single qubit flip conditioned on
the classical information obtained from the detectors in the
module $S$. In \appref{ecc}, we analyze in detail an
ECC~\cite{PerryBook} for correcting either a single bit-flip or
single phase-flip error. In this ECC, a type-II 4QM can be used
for three purposes: to implement encoding $A$ and decoding $A'$,
but also to replace two CNOT gates in the error-syndrome-detection
module $S$ (see the modules depicted in yellow in
Fig.~\ref{ECC1}). We note that CNOT-based ECCs (including the
double-controlled NOT gate, i.e., the Toffoli gate) were first
implemented experimentally in a liquid nuclear-magnetic resonance
(NMR) system~\cite{Cory98} and, later, with trapped
ions~\cite{Chiaverini04,Schindler11}, linear
optics~\cite{Pittman05}, homogenous~\cite{Moussa11} and
heterogeneous~\cite{Waldherr14} solid-state spin systems, and a
circuit-QED system~\cite{Reed12, Rist15, Kelly15, Takita16}.
Usually, the CNOT and Toffoli gates are realized by long sequences
of pulses or using qudits instead of
qubits~\cite{Reed12,Lanyon08}. In our proposal, the total number
of eight CNOT gates in the five-qubit ECC is reduced from eight to
only two, and we are not using the Toffoli gate.

Finally, we note that any entangling operation (like the type-II
4QM) is universal for quantum computing~\cite{NielsenBook}.
Specifically, by using many copies of such a gate together with
single-qubit operations, an arbitrary quantum algorithm (i.e., not
only the ECC) can be performed. However, here we focus on a direct
and simple application of the spontaneous evolution of the Dicke
system. Thus, we do not express the two remaining CNOT gates for
the syndrome detection $S$ via a sequence of 4QM and single qubit
operations.

\section{Conclusions}
In this article, we described nonlinear optical processes with
qubits, where only virtual photons are involved. The results
presented here show that $N$ spatially-separated and nondegenerate
qubits can coherently exchange energy in analogy with light modes
in nonlinear optics. The energy exchange is also reversible. This
few-body interaction is mediated by the exchange of virtual rather
than real photons and is protected against photon losses in the
bus. The coherent coupling between the $N$  superconducting qubits
can be switched on or off by tuning the transition energy of one
of them. These results can be regarded as the generalization to $N
>2$ qubits of the two-qubit coherent state transfer mediated by
virtual photons experimentally demonstrated with superconducting
artificial atoms~\cite{Majer2007}.

These processes can produce multiparticle entanglement simply
starting from one or more qubits in their excited state and
letting the system  to evolve spontaneously. The spontaneous time
evolution is also able to transfer the entanglement from a pair of
qubits to a different one. The processes proposed here extend
further the broad field of nonlinear optics. This architecture can
be extended to consider qubits in different coupled cavities and
opens up new possibilities for quantum information processing on a
chip. As an example, we have demonstrated that four-qubit mixing
can be used as a replacement of the standard quantum repetition
codes based on CNOT gates. Then, we have discussed a practical
application of type-II four-qubit mixing for quantum-information
processing, i.e., an error-correction code for encoding, decoding,
and error-syndrome detection, as shown in Fig.~\ref{ECC2} and
discussed in detail in~\appref{ecc}. Finally, we observe that
these effective three- and four-body interactions can give rise to
exotic phases not seen in usual condensed-matter
experiments~\cite{Wen2004}.

\acknowledgements We thank Simon Devitt for useful comments.
A.F.K. acknowledges support from a JSPS Postdoctoral Fellowship
for Overseas Researchers. A.M. and F.N. acknowledge the support of
a grant from the John Templeton Foundation. F.N. was also
partially supported by the RIKEN iTHES Project, the MURI Center
for Dynamic Magneto-Optics via the AFOSR award number
FA9550-14-1-0040, the IMPACT program of JST, CREST, and a
Grant-in-Aid for Scientific Research (A). S.S. acknowledge the
partial support of MPNS COST Action MP1403 Nanoscale Quantum
Optics.

\appendix

\section{Energy levels and Dynamics for Type-II four qubit mixing}
\label{S1}

\begin{figure}
    \centering
    \includegraphics[scale=0.57]{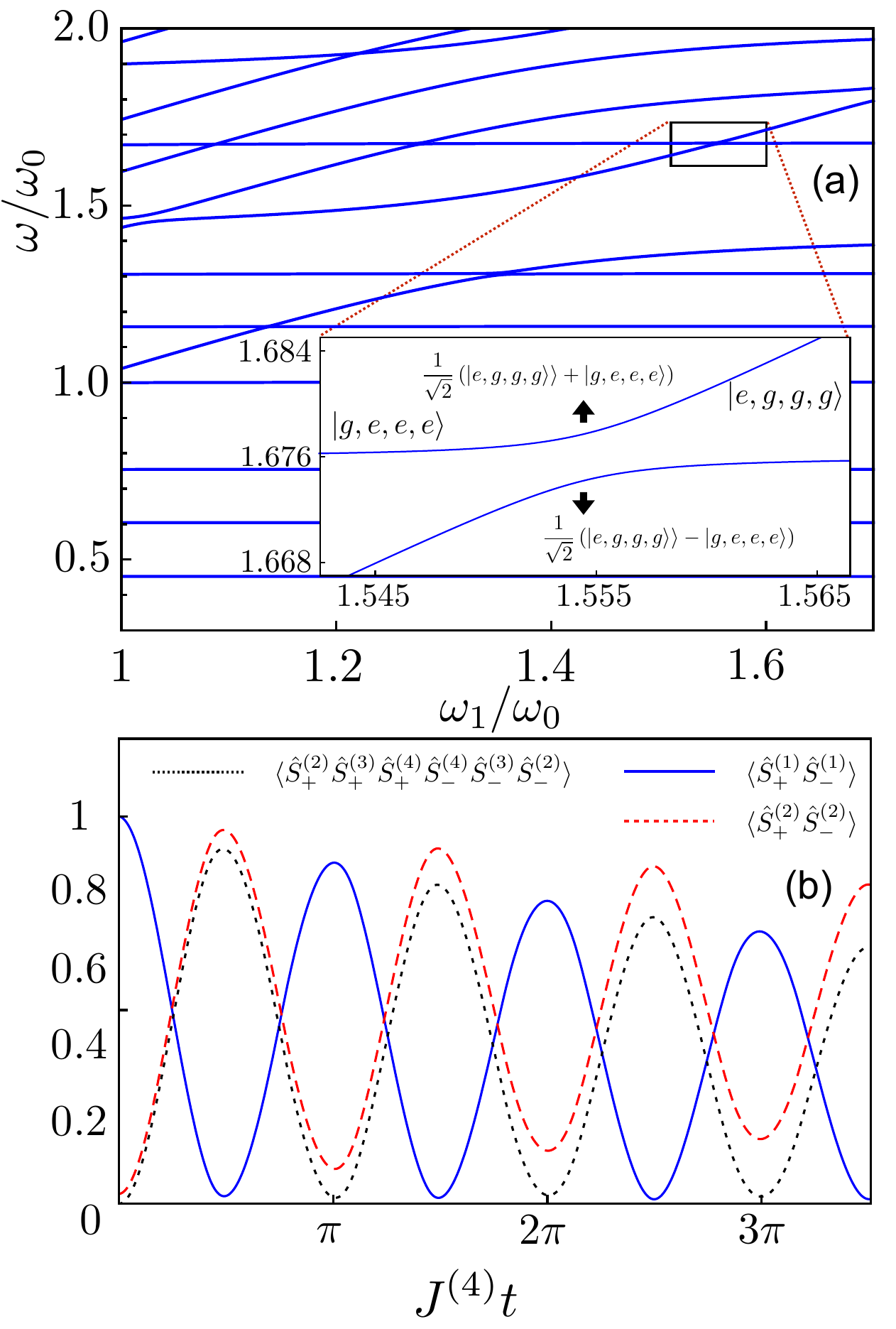}
    \caption{Type-II 4QM. (a) Lowest-energy levels of the system constituted by four qubits interacting with a cavity mode as a function of the normalized frequency of qubit 1, obtained by numerically diagonalizing the Hamiltonian in Eq.~(\ref{H0}). We used the following parameters in units of $\omega_0$: $\omega_2 = 0.4$, $\omega_3 =0.55$, $\omega_4 = 0.7$, $\omega_c= 1.75$, $\lambda_1 = 0.05$, $\lambda_i = 0.15$ (for $i = 2-4$) and $\theta_i = \pi/6$ for all the four qubits. The enlarged view of the boxed {\em apparent} crossing in the inset displays a clear anticrossing level splitting. When  the splitting is at its minimum, the eigenstates of the system are approximately  symmetric and antisymmetric superpositions of the states $|e,g,g,g,0 \rangle$ and $|g,e,e,e,0 \rangle$. (b)  Time evolution of the mean excitation probabilities of qubits 1 and 2, and of the three-qubit correlation function $\langle \hat S_+^{(2)} \hat S_+^{(3)} \hat S^{(4)}_+ \hat S^{(4)}_-\hat S^{(3)}_- \hat S^{(2)}_- \rangle$, obtained fixing $\omega_1 = 1.6448$, so that the splitting in the inset is minimum. The initial state is $|e,g,g,g,0 \rangle$ and we used $\gamma = \kappa = 3 \times 10^{-5} \omega_0$.
        \label{fig:S2}}
\end{figure}
Figure~\ref{fig:S2} presents the numerical calculations for
type-II 4QM. Figure~\ref{fig:S2}(a) displays the lowest-energy
levels as a function of the normalized frequency of qubit 1,
obtained by numerically diagonalizing the Hamiltonian in
Eq.~(\ref{H0}). Parameters are provided in the figure caption.
Panel~\ref{fig:S2}(a) also shows the enlarged view of the boxed
{\em apparent} crossing. Figure~\ref{fig:S2}(b) shows the time
evolution of the mean excitation probabilities of qubits 1 and 2,
and of the three-qubit correlation function $\langle \hat
S_+^{(2)} \hat S_+^{(3)} \hat S^{(4)}_+ \hat S^{(4)}_-\hat
S^{(3)}_- \hat S^{(2)}_- \rangle$, obtained fixing $\omega_1 =
1.6448$, so that the splitting in the inset in (a) is minimum. The
initial state is $|e,g,g,g,0 \rangle$.


\section
{Analytical derivation of the three-qubit coupling strength
$J^{(3)}$} \label{app:3Qubits} 
\begin{figure*}
    \centering
    \includegraphics[width= 16 cm]{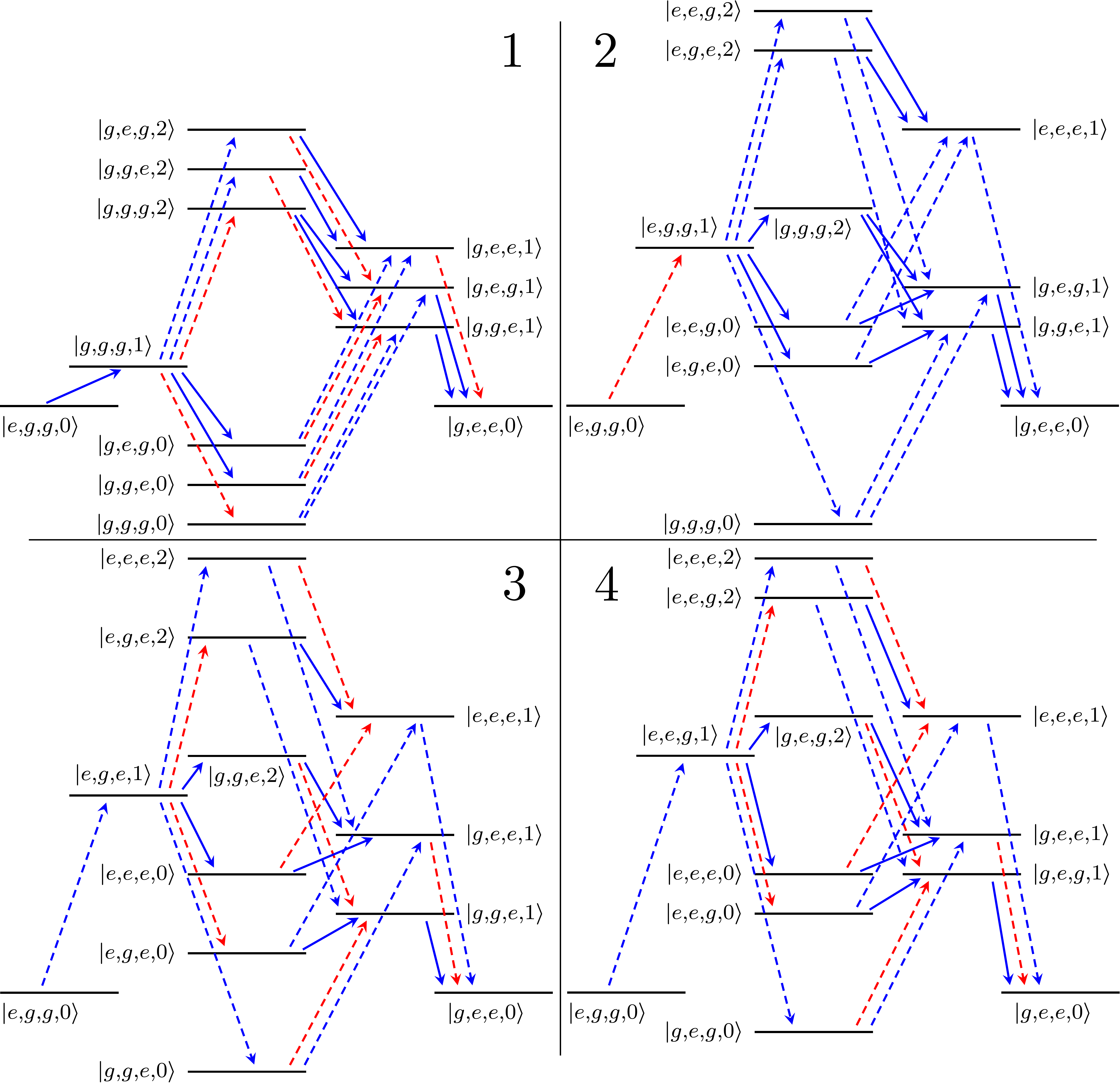}
    \caption{The 48 fourth-order paths connecting the states $\ket{e,g,g,0}$ and $\ket{g,e,e,0}$. The four diagrams show 12 paths each. Diagram 1 shows the paths starting with $\ket{e,g,g,0} \to \ket{g,g,g,1}$, diagram 2 shows the paths starting with $\ket{e,g,g,0} \to \ket{e,g,g,1}$, diagram 3 shows the paths starting with $\ket{e,g,g,0} \to \ket{e,g,e,1}$, and diagram 4 shows the paths starting with $\ket{e,g,g,0} \to \ket{e,e,g,1}$. Transitions that do not conserve the number of excitations in the system are marked by dashed lines, while transitions that conserve the number of excitations are marked by solid lines. The red lines mark the transitions mediated by the $\sz$ part of the coupling and the blue lines mark the transitions mediated by the $\sx$ part of the coupling. To set the energy levels, we have used the parameter values $\omega_c = 4\omega_3$, $\omega_1 = 3\omega_3$, and $\omega_2 = 2\omega_3$. \label{fig:48Paths}}
\end{figure*}

Our system consists of three
qubits, all ultrastrongly coupled to a cavity mode. The
Hamiltonian for the system is given in \eqref{H0}. In this
Appendix, we calculate the effective 3QM coupling strength
$J^{(3)}$ between the two states $\ket{i} = \ket{e,g,g,0}$ and
$\ket{f} = \ket{g,e,e,0}$ (equivalent to the 3QM $\ket{g,g,e,0}
\leftrightarrow \ket{e,e,g,0}$, discussed in the main text, with a
permutation of the qubit indices). These states are connected via
48 fourth-order paths, as shown in \figref{fig:48Paths}.

We can treat the interaction part of the Hamiltonian, given in
\eqref{eq:V3QM}, as a perturbation. As discussed in the main text,
the effective coupling is then given by \eqref{J3}, where the sum
goes over all paths shown in \figref{fig:48Paths}. There are no
contributions from terms of order lower than four. Note that $E_i
= E_f = 0$ when $\omega_1 = \omega_2 + \omega_3$, which is the
case considered here.

In the rest of this Appendix, for brevity, we use the notation
$\Delta_{nm} = \omega_n - \omega_m$, $\Delta_{Cn} = 2\omega_c -
\omega_n$, $\Omega_{nm} = \omega_n + \omega_m$, $\Omega_{Cn} =
2\omega_c + \omega_n$, $\lambda_{\pm\pm\pm} = \pm \lambda_1 \pm
\lambda_2 \pm \lambda_3$, $\Lambda_3 = \lambda_1 \lambda_2
\lambda_3$, and $\Lambda_4 = \lambda_1 \lambda_2 \lambda_3
\lambda_4$. With this notation, the contribution from the terms in
diagram 1 becomes
\bea
\lambda_{\rm eff}^{(1)} &=& - \frac{\Lambda_3 \sin \theta \cos^3 \theta}{\Delta_{c1}} \bigg[ 2\frac{\lambda_{-++}}{\omega_c \Delta_{C3}} + 2\frac{\lambda_{-+-}}{\Delta_{C2} \Delta_{c3}} \nn\\
&&+ 2\frac{\lambda_{-++}}{\omega_c \Delta_{C2}} + 2\frac{\lambda_{--+}}{\Delta_{C2} \Delta_{c2}} + 2\frac{\lambda_{---}}{\Delta_{C1} \Delta_{c3}} + 2\frac{\lambda_{---}}{\Delta_{C1} \Delta_{c1}} \nn\\
&&+ \frac{\lambda_{-++}}{\omega_c \left(- \omega_3 \right)} +
\frac{\lambda_{-+-}}{\left(- \omega_1 \right) \Delta_{c1}}+
\frac{\lambda_{-++}}{\omega_c \left(- \omega_2 \right)}
+ \frac{\lambda_{--+}}{\left(- \omega_2 \right) \Delta_{c2}}\nn\\
&&+ \frac{\lambda_{---}}{\left(- \omega_1 \right) \Delta_{c3}} +
\frac{\lambda_{---}}{\left(- \omega_1 \right) \Delta_{c2}} \bigg].
\label{eq:geff1}
\eea %

To check our calculations, we can compare with the calculation in
Ref.~\cite{Garziano2016}, which treated the process $\ket{g,g,1}
\to \ket{e,e,0}$. This is exactly the process in diagram 1 after
the first transition, ignoring the first qubit. If we insert the
values $\omega_c = 2\omega_q$, $\omega_2 = \omega_3 = \omega_0$,
$\omega_1 = 2\omega_0$ [recall that we have used the fact that
$\omega_1 = \omega_2 + \omega_3$ to derive \eqref{eq:geff1}],
$\lambda_2 = \lambda_3 = \lambda$, and $\lambda_1 = 0$ in
\eqref{eq:geff1}, and remove the factor
$-\lambda_1\cos\theta/\Delta_{c1}$ (which comes from the first
transition in the diagram), we obtain
\be \lambda_{\rm eff}^{\rm (1ph,2qb)} = -\frac{8}{3}\sin\theta
\cos^2\theta \frac{\lambda^3}{\omega_0^2}, \ee
which is exactly the result from Ref.~\cite{Garziano2016}.

If we make the simplifying assumptions $\lambda_1 = \lambda_2 =
\lambda_3 \equiv \lambda$, $\omega_1 \equiv \omega_0$, and
$\omega_2 = \omega_3 = \omega_0/2$, the contribution from diagram
1 to the coupling becomes
\bea
\lambda_{\rm eff}^{(1)} &=& - \frac{2\lambda^4 \sin\theta \cos^3\theta}{\omega_c - \omega_0} \bigg[ \frac{2}{\omega_c \left(2\omega_c - \frac{1}{2} \omega_0 \right)} \nn\\
&&- \frac{2}{\left(\omega_c - \frac{1}{2} \omega_0 \right) \left(2\omega_c - \frac{1}{2} \omega_0 \right)} - \frac{12}{\left(2\omega_c - \omega_0 \right)^2} \nn\\
&& - \frac{2}{\omega_c \omega_0} + \frac{5}{\omega_0 \left(
\omega_c - \frac{1}{2} \omega_0 \right)} \bigg].
\eea
%

The terms from diagram 2 add up to
\bea
\lambda_{\rm eff}^{(2)} &=& - \frac{\Lambda_3 \lambda_{+--} \sin\theta \cos^3 \theta}{\omega_c} \bigg[ \frac{2}{\Omega_{C2} \Omega_{c1}} + \frac{2}{\Omega_{C2} \Delta_{c3}} \nn\\
&&+ \frac{2}{\Omega_{C3} \Omega_{c1}} + \frac{2}{\Omega_{C3} \Delta_{c2}} + \frac{2}{\Delta_{C1} \Delta_{c3}} \nn\\
&&+ \frac{2}{\Delta_{C1} \Delta_{c2}} + \frac{1}{\omega_2 \Omega_{c1}}+ \frac{1}{\omega_2 \Delta_{c3}} + \frac{1}{\omega_3 \Omega_{c1}} \nn\\
&&+ \frac{1}{\omega_3 \Delta_{c2}} + \frac{1}{\left(-
\omega_1\right) \Delta_{c3}} + \frac{1}{\left(- \omega_1\right)
\Delta_{c2}} \bigg]. \eea
and with the same assumptions as above ($\lambda_1 = \lambda_2 =
\lambda_3 \equiv \lambda$, $\omega_1 \equiv \omega_0$, and
$\omega_2 = \omega_3 = \omega_0/2$) this simplifies to
\bea
\lambda_{\rm eff}^{(2)} &=& \frac{2\lambda^4 \sin\theta \cos^3 \theta}{\omega_c} \bigg[ \frac{2}{\left(2\omega_c + \frac{1}{2}\omega_0 \right) \left(\omega_c + \omega_0 \right)}\nn\\
&&+ \frac{2}{\left(2\omega_c + \frac{1}{2}\omega_0 \right) \left(\omega_c - \frac{1}{2}\omega_0 \right)} + \frac{4}{\left(2\omega_c - \omega_0 \right)^2} \nn\\
&&+ \frac{2}{\omega_0  \left(\omega_c + \omega_0 \right)} +
\frac{1}{\omega_0  \left(\omega_c - \frac{1}{2} \omega_0 \right)}
\bigg]. \eea

The terms from diagram 3 add up to
\bea
\lambda_{\rm eff}^{(3)} &=& - \frac{\Lambda_3 \sin\theta \cos^3 \theta}{\Omega_{c3}} \bigg[ 2\frac{\lambda_{+++}}{\Omega_{C1} \Omega_{c1}} + 2\frac{\lambda_{-++}}{\Omega_{C1} \omega_c} + 2\frac{\lambda_{+-+}}{\Omega_{C3} \Omega_{c1}} \nn\\
&&+ 2\frac{\lambda_{+-+}}{\Omega_{C3} \Delta_{c2}} + 2\frac{\lambda_{-++}}{\Delta_{C2} \omega_c} + 2\frac{\lambda_{--+}}{\Delta_{C2} \Delta_{c2}} + \frac{\lambda_{+++}}{\omega_1 \Omega_{c1}} \nn\\
&&+ \frac{\lambda_{-++}}{ \omega_1 \omega_c} + \frac{\lambda_{+-+}}{\omega_3 \Omega_{c1}} + \frac{\lambda_{+-+}}{\omega_3 \Delta_{c2}}+ \frac{\lambda_{-++}}{\left( - \omega_2 \right) \omega_c} \nn\\
&&+ \frac{\lambda_{--+}}{\left( - \omega_2 \right) \Delta_{c2}}
\bigg], \label{eq:Geff3} \eea
and with the same assumptions as above this simplifies to
\begin{widetext}
\bea
\lambda_{\rm eff}^{(3)} &=& -\frac{\lambda^4\sin\theta\cos^3\theta}{\omega_c + \frac{1}{2}\omega_0} \bigg[ \frac{6}{\left(2\omega_c + \omega_0 \right) \left(\omega_c + \omega_0 \right)} + \frac{2}{\left(2\omega_c + \omega_0 \right) \omega_c} + \frac{2}{\left(2\omega_c + \frac{1}{2}\omega_0 \right) \left(\omega_c + \omega_0 \right)} \nn\\
&& + \frac{2}{\left(2\omega_c + \frac{1}{2}\omega_0 \right) \left(\omega_c - \frac{1}{2}\omega_0 \right)}+ \frac{2}{\left(2\omega_c - \frac{1}{2}\omega_0 \right) \omega_c} - \frac{2}{\left(2\omega_c - \frac{1}{2}\omega_0 \right) \left(\omega_c - \frac{1}{2}\omega_0 \right)}  \nn\\
&& + \frac{5}{\omega_0 \left(\omega_c + \omega_0 \right)} -
\frac{1}{\omega_0\omega_c} + \frac{4}{\omega_0 \left(\omega_c -
\frac{1}{2}\omega_0 \right)} \bigg]. \label{eq:GeffSimp3} \eea
%

The terms from diagram 4 add up to $\lambda_{\rm eff}^{(4)}$,
which is given by \eqref{eq:Geff3} but with the indices 2 and 3
interchanged everywhere. With the same assumptions as above
$\lambda_{\rm eff}^{(4)}$ simplifies to exactly the same
expression as for diagram 3, given in \eqref{eq:GeffSimp3}. Adding
up all the terms from the four diagrams in the simplified case
gives
\bea
\lambda_{\text{eff}} &=& -2\lambda^4 \sin\theta \cos^3\theta \bigg\{ \frac{1}{\omega_c - \omega_0} \bigg[ \frac{2}{\omega_c \left(2\omega_c - \frac{1}{2}\omega_0 \right)} - \frac{2}{\left(2\omega_c - \frac{1}{2}\omega_0 \right) \left(\omega_c - \frac{1}{2}\omega_0 \right)} - \frac{12}{\left(2\omega_c - \omega_0 \right)^2}  \nn\\
&&- \frac{2}{\omega_c\omega_0}+ \frac{5}{\left(\omega_c - \frac{1}{2}\omega_0 \right)\omega_0} \bigg] -\frac{1}{\omega_c} \bigg[ \frac{2}{\left(2\omega_c + \frac{1}{2}\omega_0 \right) \left(\omega_c + \omega_q \right)} + \frac{2}{\left(2\omega_c + \frac{1}{2}\omega_0 \right) \left(\omega_c - \frac{1}{2}\omega_0 \right)} \nn\\
&&+ \frac{4}{\left(2\omega_c - \omega_0 \right)^2}+
\frac{2}{\omega_0  \left(\omega_c + \omega_0 \right)} +
\frac{1}{\omega_0  \left(\omega_c - \frac{1}{2} \omega_0 \right)}
\bigg] + \frac{1}{\omega_c + \frac{1}{2}\omega_0} \bigg[ \frac{6}{\left(2\omega_c + \omega_0 \right) \left(\omega_c + \omega_0 \right)} \nn\\
&& + \frac{2}{\left(2\omega_c + \omega_0 \right) \omega_c}+ \frac{2}{\left(2\omega_c + \frac{1}{2}\omega_0 \right) \left(\omega_c + \omega_0 \right)} + \frac{2}{\left(2\omega_c + \frac{1}{2}\omega_0 \right) \left(\omega_c - \frac{1}{2}\omega_0 \right)} + \frac{2}{\left(2\omega_c - \frac{1}{2}\omega_0 \right) \omega_c} \nn\\
&& - \frac{2}{\left(2\omega_c - \frac{1}{2}\omega_0 \right)
\left(\omega_c - \frac{1}{2}\omega_0 \right)}+ \frac{5}{\omega_q
\left(\omega_c + \omega_0 \right)} - \frac{1}{\omega_0\omega_c} +
\frac{4}{\omega_0 \left(\omega_c - \frac{1}{2}\omega_0 \right)}
\bigg] \bigg\}\, . \label{eq:GeffSimpTot} \eea
\end{widetext}
\noindent We note that the part from diagram 1 gives the largest
contribution, since the first transition in that diagram has a
smaller energy difference than the first transitions in the other
diagrams, resulting in a smaller denominator.

The expression in \eqref{eq:GeffSimpTot} turns out to simplify
much further when everything is put on a common denominator. The
final result is
\be \lambda_{\text{eff}} =J^{(3)} = \frac{64 \lambda^4 \omega_c^2
\left(4\omega_c^2 - 7\omega_0^2 \right) \sin \theta \cos^3
\theta}{\omega_0\left(\omega_0^2 - \omega_c^2
\right)\left(\omega_0^2 - 4\omega_c^2 \right)^2},
\label{eq:GeffSimpCDApp} \ee
which is discussed further in the main text.

We note that the coupling, also in the unsimplified case, is
proportional to $\sin\theta \cos^3\theta$. Differentiating this
with respect to $\theta$ and setting the derivative to zero, we
obtain
\be 0 = \cos^2 \theta \left(\cos^2 \theta - 3 \sin^2 \theta
\right), \label{eq:LambdaThetaDerivative} \ee
Since $\cos \theta = 0$ implies $\lambda_{\text{eff}} = 0$,
\eqref{eq:LambdaThetaDerivative} gives that the maximum coupling
is achieved when
\be \theta = \pm \arctan \frac{1}{\sqrt{3}} = \pm \frac{\pi}{6}.
\ee

\section{Analytical calculations with four qubits}
\label{analqubits} Our system now consists of four qubits, all
ultrastrongly coupled to a cavity mode. We calculate the effective
coupling strength between the two states $\ket{i} =
\ket{e,e,g,g,0}$ and $\ket{f} = \ket{g,g,e,e,0}$ (equivalent to
the 4QM $\ket{e,g,g,e,0} \leftrightarrow \ket{g,e,e,g,0}$,
discussed in the main text, with a permutation of the qubit
indices). Since this transition conserves the total number of
excitations in the system, we limit our study to the case where
$\theta_i = 0$ for all four qubits in the interaction Hamiltonian
\eqref{eq:V3QM}. The Hamiltonian we work with is thus
\be \hat H = \omega_c \hat a^\dag \hat a + \sum_{i=1}^4
\frac{\omega_{i}}{2} \sz^{(i)} + \left( \hat a + \hat a^\dag
\right) \sum_{i=1}^4 \lambda_i \sx^{(i)}. \label{eq:4qbHRabi} \ee
In fact, since the number of excitations is conserved, we can also
consider only the TC terms in \eqref{eq:4qbHRabi}, i.e., the TC
Hamiltonian
\be \hat H = \omega_c \hat a^\dag \hat a + \sum_{i=1}^4
\frac{\omega_{i}}{2} \sz^{(i)} + \sum_{i=1}^4 \lambda_j \left(
\hat a \sp^{(i)} + \hat a^\dag \sm^{(i)} \right).
\label{eq:4qbHJC} \ee

\begin{figure*}
    \centerline{
        \resizebox{12 cm}{!}{
            \begin{tikzpicture}[
            scale=1.3,
            level/.style={thick},
            ztrans/.style={thick,->,shorten >=0.1cm,shorten <=0.1cm,>=stealth,densely dashed,color=red},
            nrtrans/.style={thick,->,shorten >=0.1cm,shorten <=0.1cm,>=stealth,densely dashed,color=blue},
            rtrans/.style={thick,->,shorten >=0.1cm,shorten <=0.1cm,>=stealth,color=blue},
            classical/.style={thin,double,<->,shorten >=4pt,shorten <=4pt,>=stealth},
            ]
            \coordinate (0eegg) at (0, 0); 
            \coordinate (v) at (0, 0.3); 
            \coordinate (l) at (1.2, 0); 
            \coordinate (h) at (0.4, 0); 
            \coordinate (1eggg) at ($(0eegg) + (0,0) + (l) + (h) + 5*(v)$);
            \coordinate (1gegg) at ($(0eegg) + (0,0) + (l) + (h) + 2*(v)$);
            \coordinate (2gggg) at ($(0eegg) + 2*(l) + 2*(h) + 7*(v)$);
            \coordinate (0egeg) at ($(0eegg) + 2*(l) + 2*(h) + 2*(v)$);
            \coordinate (0egge) at ($(0eegg) + 2*(l) + 2*(h) + 1*(v)$);
            \coordinate (0geeg) at ($(0eegg) + 2*(l) + 2*(h) - 1*(v)$);
            \coordinate (0gege) at ($(0eegg) + 2*(l) + 2*(h) - 2*(v)$);
            \coordinate (1ggeg) at ($(0eegg) + 3*(l) + 3*(h) + 4*(v)$);
            \coordinate (1ggge) at ($(0eegg) + 3*(l) + 3*(h) + 3*(v)$);
            \coordinate (0ggee) at ($(0eegg) + 4*(l) + 4*(h) - (0.5,0)$);
            \draw[level] ($(0eegg) + (0.5,0)$) -- ($(0eegg) + (0.5,0) + (l)$) node[midway,below,xshift=-0.3cm] {\footnotesize{$\ket{e,e,g,g,0}$}};
            \draw[level] (1eggg) -- ($(1eggg) + (l)$) node[midway,left,xshift=-0.7cm] {\footnotesize{$\ket{e,g,g,g,1}$}};
            \draw[level] (1gegg) -- ($(1gegg) + (l)$) node[midway,left,xshift=-0.7cm] {\footnotesize{$\ket{g,e,g,g,1}$}};
            \draw[level] (2gggg) -- ($(2gggg) + (l)$) node[midway,above,xshift=0cm] {\footnotesize{$\ket{g,g,g,g,2}$}};
            \draw[level] (0egeg) -- ($(0egeg) + (l)$) node[midway,above,yshift=0.05cm,xshift=0cm] {\footnotesize{$\ket{e,g,e,g,0}$}};
            \draw[level] (0egge) -- ($(0egge) + (l)$) node[midway,below,yshift=0.05cm,xshift=0cm] {\footnotesize{$\ket{e,g,g,e,0}$}};
            \draw[level] (0geeg) -- ($(0geeg) + (l)$) node[midway,left,xshift=-0.8cm] {\footnotesize{$\ket{g,e,e,g,0}$}};
            \draw[level] (0gege) -- ($(0gege) + (l)$) node[midway,left,xshift=-0.8cm] {\footnotesize{$\ket{g,e,g,e,0}$}};
            \draw[level] (1ggeg) -- ($(1ggeg) + (l)$) node[midway,right,xshift=0.7cm] {\footnotesize{$\ket{g,g,e,g,1}$}};
            \draw[level] (1ggge) -- ($(1ggge) + (l)$) node[midway,right,xshift=0.7cm] {\footnotesize{$\ket{g,g,g,e,1}$}};
            \draw[level] (0ggee) -- ($(0ggee) + (l)$) node[midway,below,xshift=0cm] {\footnotesize{$\ket{g,g,e,e,0}$}};
            \draw[rtrans] ($(0eegg) + (1.5cm,0cm)$) -- ($(1eggg) + (0.6cm,0cm)$);
            \draw[rtrans] ($(0eegg) + (1.6cm,0cm)$) -- ($(1gegg) + (0.5cm,0cm)$);
            \draw[rtrans] ($(1eggg) + (1.0cm,0cm)$) -- ($(2gggg) + (0.35cm,0cm)$);
            \draw[rtrans] ($(1eggg) + (0.85cm,0cm)$) -- ($(0egeg) + (0.0cm,0cm)$);
            \draw[rtrans] ($(1eggg) + (0.7cm,0cm)$) -- ($(0egge) + (0.1cm,0cm)$);
            \draw[rtrans] ($(1gegg) + (1.0cm,0cm)$) -- ($(2gggg) + (0.5cm,0cm)$);
            \draw[rtrans] ($(1gegg) + (1.15cm,0cm)$) -- ($(0geeg) + (0.05cm,0cm)$);
            \draw[rtrans] ($(1gegg) + (1.05cm,0cm)$) -- ($(0gege) + (0.1cm,0cm)$);
            \draw[rtrans] ($(2gggg) + (1.1cm,0cm)$) -- ($(1ggeg) + (0.3cm,0cm)$);
            \draw[rtrans] ($(2gggg) + (1.0cm,0cm)$) -- ($(1ggge) + (0.35cm,0cm)$);
            \draw[rtrans] ($(0egeg) + (1.1cm,0cm)$) -- ($(1ggeg) + (0.1cm,0cm)$);
            \draw[rtrans] ($(0egge) + (1.1cm,0cm)$) -- ($(1ggge) + (0.15cm,0cm)$);
            \draw[rtrans] ($(0geeg) + (1.1cm,0cm)$) -- ($(1ggeg) + (0.6cm,0cm)$);
            \draw[rtrans] ($(0gege) + (1.1cm,0cm)$) -- ($(1ggge) + (0.65cm,0cm)$);
            \draw[rtrans] ($(1ggeg) + (0.8cm,0cm)$) -- ($(0ggee) + (0.3cm,0cm)$);
            \draw[rtrans] ($(1ggge) + (0.75cm,0cm)$) -- ($(0ggee) + (0.15cm,0cm)$);
            \end{tikzpicture}
        }
    }
    \caption{The 8 fourth-order paths connecting the states $\ket{e,e,g,g,0}$ and $\ket{g,g,e,e,0}$ using only the TC terms. To set the energy levels, we have used the parameter values $\omega_c = 6\omega_2$, $\omega_1 = 4\omega_2$, $\omega_3 = 3\omega_2$, and $\omega_4 = 2\omega_2$. \label{fig:JCPaths}}
\end{figure*}
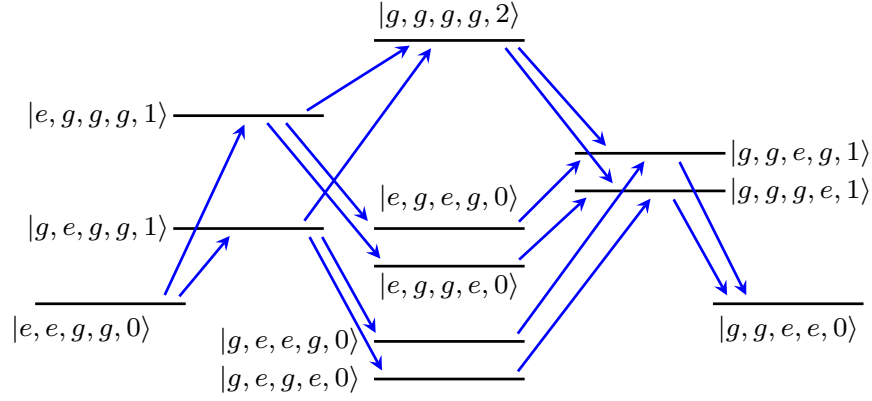

\begin{figure*}
    \centering
    \includegraphics[width=16 cm]{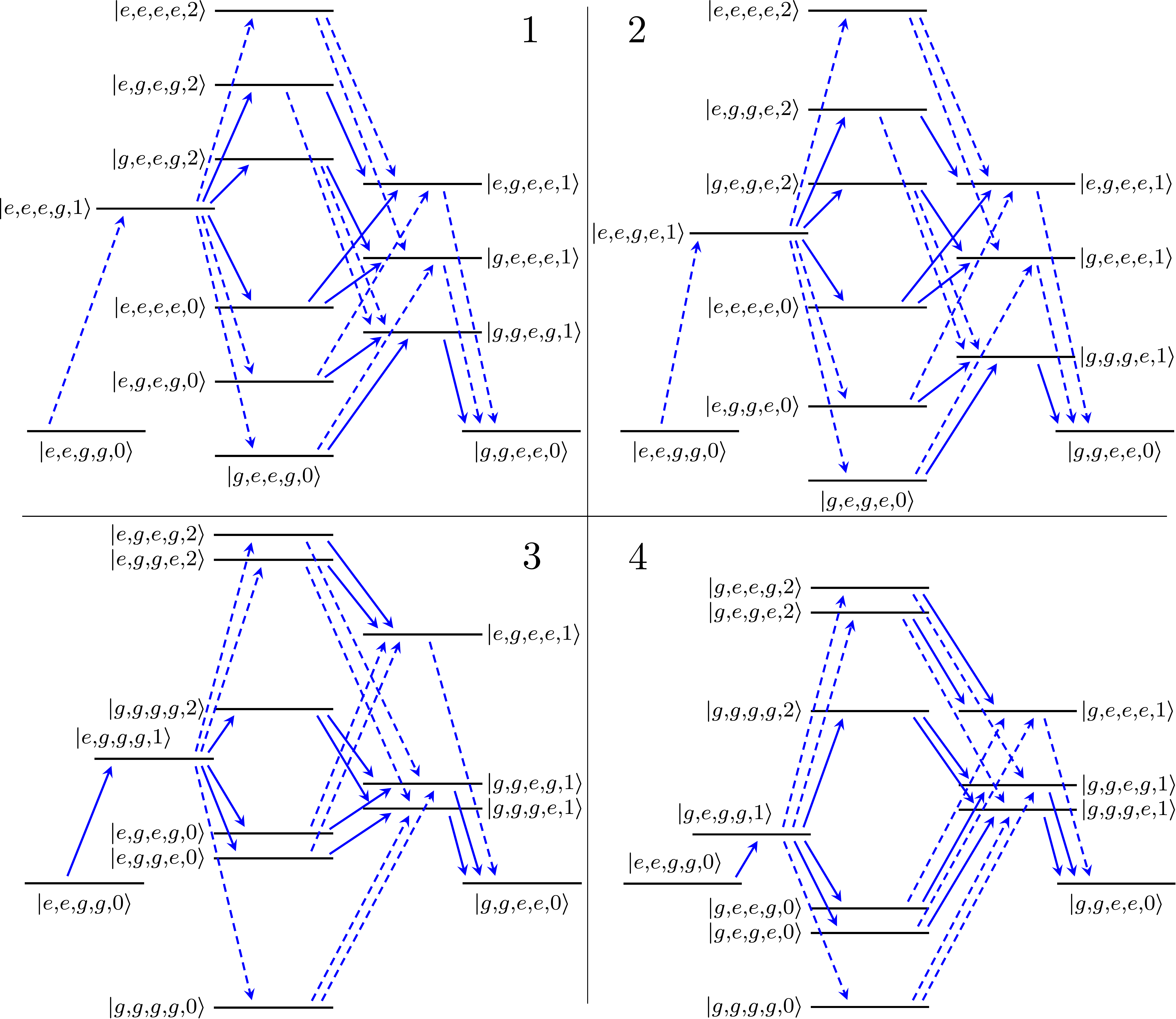}
    \caption{The 48 fourth-order paths connecting the states $\ket{e,e,g,g,0}$ and $\ket{g,g,e,e,0}$. The four diagrams show 12 paths each. Diagram 1 shows the paths starting with $\ket{e,e,g,g,0} \to \ket{e,e,e,g,1}$, diagram 2 shows the paths starting with $\ket{e,e,g,g,0} \to \ket{e,e,g,e,1}$, diagram 3 shows the paths starting with $\ket{e,e,g,g,0} \to \ket{e,g,g,g,1}$, and diagram 4 shows the paths starting with $\ket{e,e,g,g,0} \to \ket{g,e,g,g,1}$. Transitions that do not conserve the number of excitations in the system are marked by dashed lines, while the transitions that conserve the number of excitations are marked by solid lines. To set the energy levels, we have used the same parameter values as in \figref{fig:JCPaths}. \label{fig:48Paths4Qubits}}
\end{figure*}

The two states $\ket{e,e,g,g,0}$ and $\ket{g,g,e,e,0}$ are
connected via fourth-order processes. If we use \eqref{eq:4qbHJC},
there are 8 paths connecting the states to this order, as shown in
\figref{fig:JCPaths}. If we use \eqref{eq:4qbHRabi} instead, there
are 48 paths connecting the states to this order, as shown in
\figref{fig:48Paths4Qubits}.

\subsection{Calculations with the Tavis-Cummings Hamiltonian}

We use the same calculation method, as well as the notation, as
in \ref{app:3Qubits}. From Eqs.~(\ref{J3}) and (\ref{eq:4qbHJC})
and \figref{fig:JCPaths}, we then obtain the effective coupling
\begin{widetext}
\bea
\lambda_{\rm eff} &=& - \Lambda_4 \bigg[ \frac{2}{\Delta_{c2} \left( \Delta_{c1} + \Delta_{c2} \right) \left( \Delta_{c1} + \Delta_{32} \right) } + \frac{2}{\Delta_{c2} \left( \Delta_{c1} + \Delta_{c2} \right) \left( \Delta_{c1} + \Delta_{42} \right) } \nn\\
&&+ \frac{1}{\Delta_{c2} \Delta_{32} \left( \Delta_{c1} + \Delta_{32} \right) } + \frac{1}{\Delta_{c2} \Delta_{42} \left( \Delta_{c1} + \Delta_{42} \right) } + \frac{2}{\Delta_{c1} \left( \Delta_{c1} + \Delta_{c2} \right) \left( \Delta_{c1} + \Delta_{32} \right) } \nn\\
&&+ \frac{2}{\Delta_{c1} \left( \Delta_{c1} + \Delta_{c2} \right) \left( \Delta_{c1} + \Delta_{42} \right) } + \frac{1}{\Delta_{c1} \Delta_{31} \left( \Delta_{c1} + \Delta_{32} \right) } + \frac{1}{\Delta_{c1} \Delta_{41} \left( \Delta_{c1} + \Delta_{42} \right) } \bigg] \nn\\
&=& \frac{\Lambda_4 \left( \Delta_{13} + \Delta_{24} \right)
\left( \Delta_{13} \Delta_{24} + \Delta_{14} \Delta_{23}
\right)}{\Delta_{13} \Delta_{23} \Delta_{14} \Delta_{24}
\Delta_{1c} \Delta_{2c}}. \eea

Because of the factor $\omega_1 + \omega_2 - \omega_3 - \omega_4$
in the numerator, this expression goes to zero on resonance
($\omega_1 + \omega_2 = \omega_3 + \omega_4$).

\subsection{Calculations with the quantum Rabi Hamiltonian}

The first diagram in \figref{fig:48Paths4Qubits} gives the
following contribution to the effective coupling:
\bea
\lambda_{\rm eff}^{(1)} &=& - \frac{\Lambda_4}{\Omega_{c3}} \bigg[ \frac{2}{\left( \Omega_{c3} + \Omega_{c4} \right) \left( \Omega_{c3} + \Delta_{42} \right) } + \frac{2}{\left( \Omega_{c3} + \Omega_{c4} \right) \left( \Omega_{c3} + \Delta_{41} \right)} \nn\\
&&+ \frac{2}{\left( \Omega_{c3} + \Delta_{c2} \right) \left( \Omega_{c3} + \Delta_{42} \right)} + \frac{2}{\left( \Omega_{c3} + \Delta_{c2} \right) \left( \Omega_{c3} - \Omega_{12} \right)} + \frac{2}{\left( \Omega_{c3} + \Delta_{c1} \right) \left( \Omega_{c3} + \Delta_{41} \right)} \nn\\
&&+ \frac{2}{\left( \Omega_{c3} + \Delta_{c1} \right) \left( \Omega_{c3} - \Omega_{12} \right)} + \frac{1}{\Omega_{34} \left( \Omega_{c3} + \Delta_{42} \right) } + \frac{1}{\Omega_{34} \left( \Omega_{c3} + \Delta_{41} \right)}  \nn\\
&& + \frac{1}{\Delta_{32} \left( \Omega_{c3} + \Delta_{41}
\right)} + \frac{1}{\Delta_{32} \left( \Omega_{c3} - \Omega_{12}
\right)} + \frac{1}{\Delta_{31} \left( \Omega_{c3} + \Delta_{41}
\right)} + \frac{1}{\Delta_{31} \left( \Omega_{c3} - \Omega_{12}
\right)} \bigg]. \eea
Note that the six last terms are just the six first terms with $2$
replaced by $1$ in the numerator and $2\omega_c$ replaced by zero
in the denominator. This holds for all four diagrams. The terms in
the second diagram contribute with
\bea
\lambda_{\rm eff}^{(2)} &=& - \frac{\Lambda_4}{\Omega_{c4}} \bigg[ \frac{2}{\left( \Omega_{c3} + \Omega_{c4} \right) \left( \Omega_{c3} + \Delta_{42} \right) } + \frac{2}{\left( \Omega_{c3} + \Omega_{c4} \right) \left( \Omega_{c3} + \Delta_{41} \right)} \nn\\
&&+ \frac{2}{\left( \Omega_{c4} + \Delta_{c2} \right) \left( \Omega_{c3} + \Delta_{42} \right)} + \frac{2}{\left( \Omega_{c4} + \Delta_{c2} \right) \left( \Omega_{c4} - \Omega_{12} \right)} + \frac{2}{\left( \Omega_{c4} + \Delta_{c1} \right) \left( \Omega_{c3} + \Delta_{41} \right)} \nn\\
&&+ \frac{2}{\left( \Omega_{c4} + \Delta_{c1} \right) \left( \Omega_{c4} - \Omega_{12} \right)} + \frac{1}{\Omega_{34} \left( \Omega_{c3} + \omega_4 - \omega_2 \right) } + \frac{1}{\Omega_{34} \left( \Omega_{c3} + \omega_4 - \omega_1 \right)}\nn\\
&&+ \frac{1}{\Delta_{42} \left( \Omega_{c3} + \Delta_{42} \right)}
+ \frac{1}{\Delta_{42} \left( \Omega_{c4} - \Omega_{12} \right)} +
\frac{1}{\Delta_{41} \left( \Omega_{c3} + \Delta_{41} \right)} +
\frac{1}{\Delta_{41} \left( \Omega_{c4} - \Omega_{12} \right)}
\bigg], \eea
the terms in the third diagram give
\bea
\lambda_{\rm eff}^{(3)} &=& - \frac{\Lambda_4}{\Delta_{c2}} \bigg[ \frac{2}{\left( \Omega_{c3} + \Delta_{c2} \right) \left( \Omega_{c3} + \Delta_{42} \right) } + \frac{2}{\left( \Omega_{c3} + \Delta_{c2} \right) \left( \Omega_{c3} - \Omega_{12} \right)}  \nn\\
&&+ \frac{2}{\left( \Omega_{c4} + \Delta_{c2} \right) \left( \Omega_{c3} + \Delta_{42} \right)} + \frac{2}{\left( \Omega_{c4} + \Delta_{c2} \right) \left( \Omega_{c4} - \Omega_{12} \right)} + \frac{2}{\left( \Delta_{c1} + \Delta_{c2} \right) \left( \Omega_{c3} - \Omega_{12} \right)} \nn\\
&&+ \frac{2}{\left( \Delta_{c1} + \Delta_{c2} \right) \left( \Omega_{c4} - \Omega_{12} \right)} + \frac{1}{\Delta_{32} \left( \Omega_{c3} + \Delta_{42} \right)} + \frac{1}{\Delta_{32} \left( \Omega_{c3} - \Omega_{12} \right)} + \frac{1}{\Delta_{42} \left( \Omega_{c3} + \Delta_{42} \right)} \nn\\
&&+ \frac{1}{\Delta_{42} \left( \Omega_{c4} - \Omega_{12} \right)}
+ \frac{1}{ - \Omega_{12} \left( \Omega_{c3} - \Omega_{12}
\right)} + \frac{1}{ - \Omega_{12} \left( \Omega_{c4} -
\Omega_{12}\right)} \bigg], \eea
and the terms in the fourth diagram give
\bea
\lambda_{\rm eff}^{(4)} &=& - \frac{\Lambda_4}{\Delta_{c1}} \bigg[ \frac{2}{\left( \Omega_{c3} + \Delta_{c1} \right) \left( \Omega_{c3} + \Delta_{41} \right) } + \frac{2}{\left( \Omega_{c3} + \Delta_{c1} \right) \left( \Omega_{c3} - \Omega_{12} \right)}\nn\\
&&+ \frac{2}{\left( \Omega_{c4} + \Delta_{c1} \right) \left( \Omega_{c3} + \Delta_{41} \right)} + \frac{2}{\left( \Omega_{c4} + \Delta_{c1} \right) \left( \Omega_{c4} - \Omega_{12} \right)} + \frac{2}{\left( \Delta_{c1} + \Delta_{c2} \right) \left( \Omega_{c3} - \Omega_{12} \right)} \nn\\
&&+ \frac{2}{\left( \Delta_{c1} + \Delta_{c2} \right) \left( \Omega_{c4} - \Omega_{12} \right)} + \frac{1}{\Delta_{31} \left( \Omega_{c3} + \Delta_{41} \right) } + \frac{1}{\Delta_{31} \left( \Omega_{c3} - \Omega_{12} \right)} + \frac{1}{\Delta_{41} \left( \Omega_{c3} + \Delta_{41} \right)} \nn\\
&&+ \frac{1}{\Delta_{41} \left( \Omega_{c4} - \Omega_{12} \right)} + \frac{1}{ - \Omega_{12} \left( \Omega_{c3} - \Omega_{12} \right)} + \frac{1}{ - \Omega_{12} \left( \Omega_{c4} - \Omega_{12} \right)} \bigg]. \nn\\
\eea
Adding up all these terms gives the complicated expression
\bea \lambda_{\rm eff} &=& \Lambda_4 \left( \Omega_{12} -
\Omega_{34} \right)
\bigg( 3 \omega_2 \omega_3 \omega_4 \Delta_{23} \Delta_{24} \Omega_{34} \nn\\
&&+ \left\{ 2 \omega_c  \left[ \Delta_{23} - \omega_4 \right] - 4 \omega_c^2 \right\}  \left\{ \omega_3^2 \omega_4^2 - 3 \omega_2 \omega_3 \omega_4 \Omega_{34} + \omega_2^2 \left[ \omega_3^2 + 3 \omega_3 \omega_4 + \omega_4^2 \right] \right\} \nn\\
&&+ \omega_1^2 \left\{ \Omega_{12} - \Omega_{34} - 2 \omega_c \right\} \left\{ 3 \Delta_{23} \Delta_{24} \Omega_{34} + 2 \omega_c \left[ \omega_2^2 + \omega_3^2 + \omega_4^2 + 3 \omega_3 \omega_4 - 3 \omega_2 \Omega_{34} \right] \right\} \nn\\
&& + \omega_1 \bigg\{ 12 \omega_c^2 \Delta_{23} \Delta_{24} \Omega_{34} - 3 \Delta_{23} \Delta_{24} \Omega_{34} \left[\omega_2 \Omega_{34} - \omega_3 \omega_4 \right] \nn\\
&&\quad + 2 \omega_c \big[ \omega_2^2 \left( 7 \omega_3^2 + 15 \omega_3 \omega_4 + 7 \omega_4^2 \right) + \omega_3 \omega_4 \left( 3 \omega_3^2 + 7 \omega_3 \omega_4 + 3 \omega_4^2 \right) \nn\\
&&\quad\quad - 3 \omega_2 \Omega_{34} \left( \omega_2^2 +
\omega_3^2 + 4 \omega_3 \omega_4 \right) \big] \bigg\} \bigg)
\bigg/ \Omega_{12} \Omega_{34} \Omega_{c3} \Omega_{c4} \Delta_{13}
\Delta_{14} \Delta_{23} \Delta_{24} \Delta_{c1} \Delta_{c2}. \eea

Because of the factor $\omega_1 + \omega_2 - \omega_3 - \omega_4$
in the numerator, this expression goes to zero when $\omega_1 +
\omega_2 = \omega_3 + \omega_4$. However, $\omega_i$ describe bare
qubit transition frequencies. Owing to the presence of the
counter-rotating terms in the atom-cavity interaction Hamiltonian,
the physical transition frequencies are somewhat different from
the bare ones. Hence at the physical resonance $\omega_1 +
\omega_2 \neq \omega_3 + \omega_4$.
\end{widetext}
\begin{figure*}
    \centering
    \includegraphics[height=7cm]{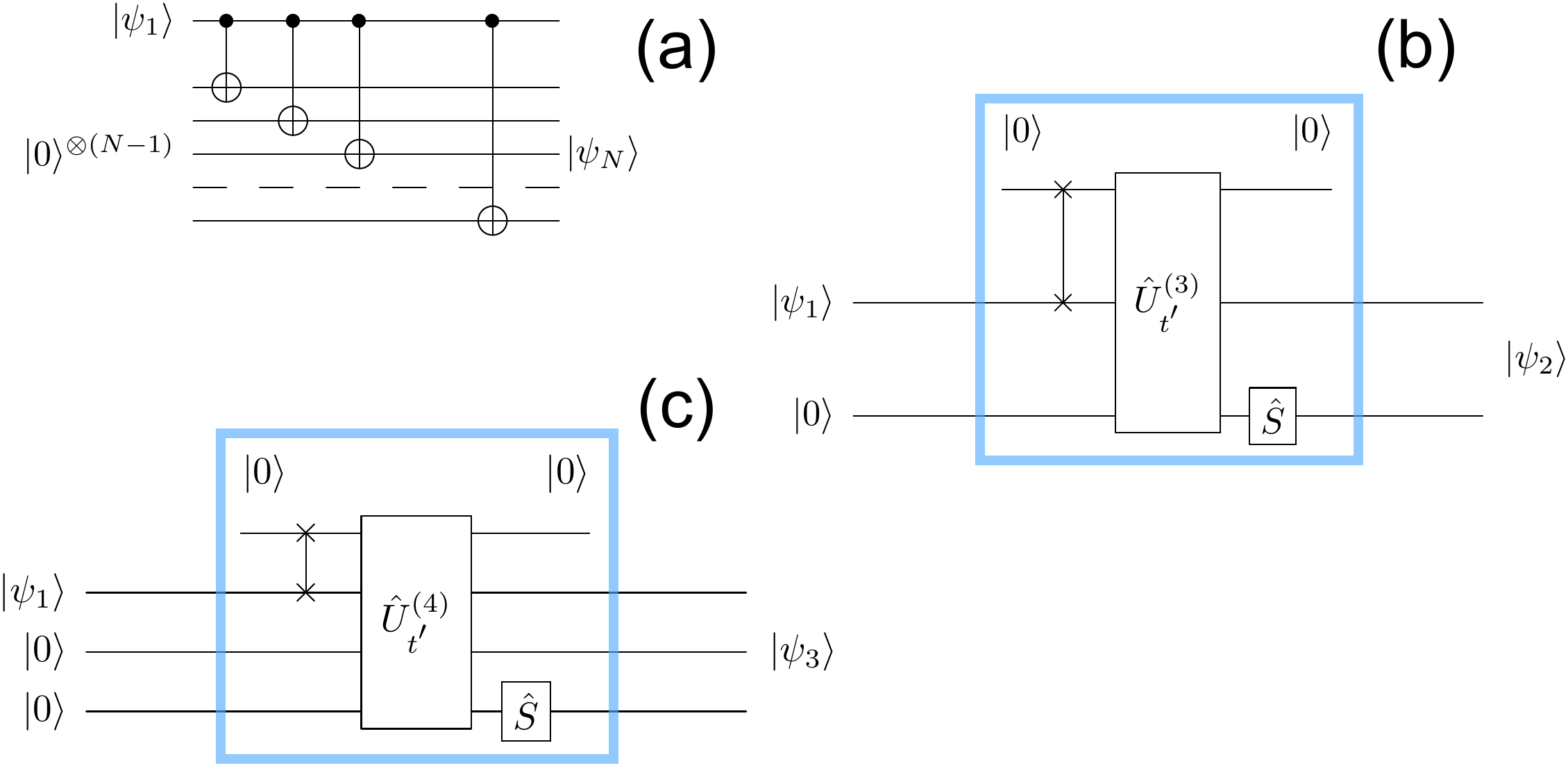}
    \caption{(a) Standard $N$-qubit repetition code based on $N$ CNOT
        gates for generating $\ket{\psi_N}=a\ket{0}^{\otimes
            N}+b\ket{1}^{\otimes N}$ (for $N=2,3,...$). Repetition codes for
        (b) three and (c) four qubits based on qubit mixing corresponding
        to the spontaneous evolution operations $\hat U^{(3)}_{t'}$ and
        $\hat U^{(4)}_{t'}$, given by Eqs.~(\ref{U3}) and~(\ref{U4}),
        respectively, followed by the $(\pi/4)$-phase gate $\hat S$ acting
        on the $N$th qubit. The operation marked by $\times\!\!
        -\!\!\times$ corresponds to the classical SWAP gate, which is not
        necessary, but added to show the correspondence of all the panels.
        In panels (b) and (c), the first qubit is finally disentangled
        from the others, so it can be discarded after the operation.  In a
        special case, the four-qubit repetition code can implement the
        encoder $A$ and decoder $A'$, and can be used for the
        error-syndrome detection in the ECC shown in Fig.~\ref{ECC2}. }
    \label{RepetitionCodesFig}
\end{figure*}

    \begin{figure*}
        \centering
        \includegraphics[width=17cm]{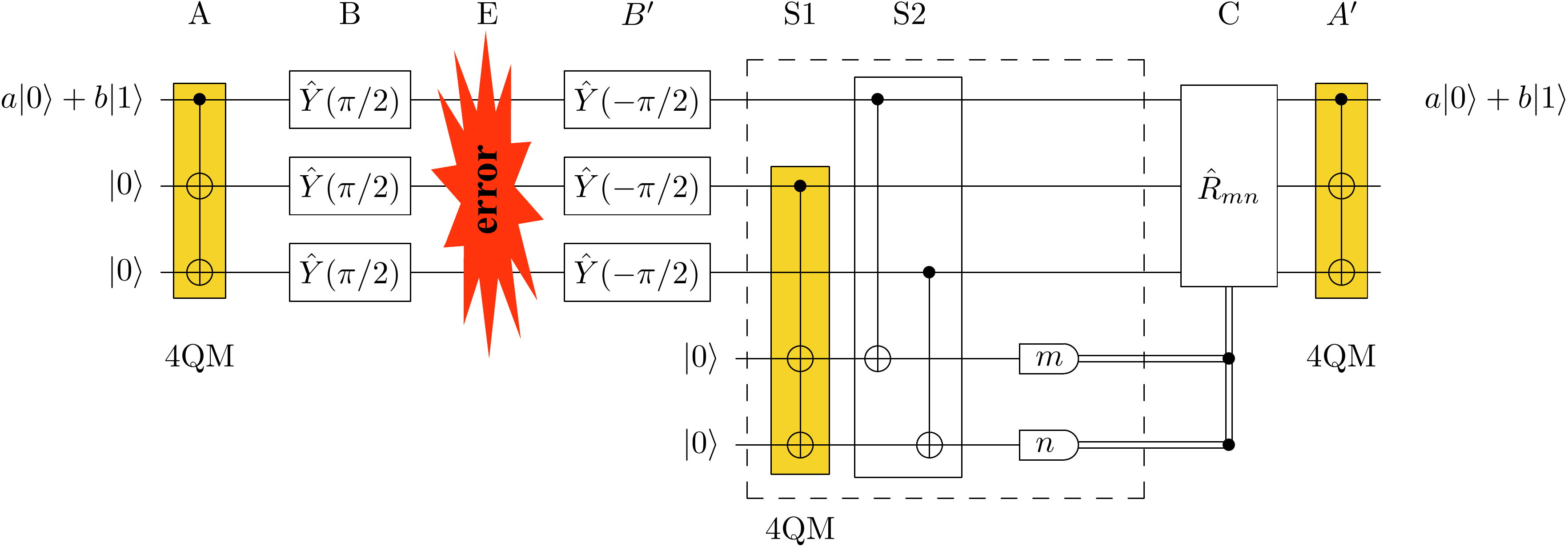}
        \caption{Circuit implementing an ECC correcting a single
            phase-flip error. If the modules $B$ and $B'$ are omitted, then
            the circuit corrects a single bit-flip error. The circuit modules
            include: encoder $A$, basis rotator $B$ (from the basis
            $\{\ket{0},\ket{1}\}$ into $\{\ket{+},\ket{-}\}$) corresponding to
            the Hadamard gates or rotations $\hat Y(\pi/2)$, dissipative
            channel or evolution ($E$, when the error happens), basis rotator
            $B'$ (from $\{\ket{+},\ket{-}\}$ into $\{\ket{0},\ket{1}\}$ being
            inverse to $B$), error-syndrome detector $S$ (composed of blocks
            $S_1$, $S_2$, and detectors with two possible outcomes $m,n=0,1$),
            error corrector ($C$, where the single-qubit operation $R_{nm}$,
            given by Eq.~(\ref{Rmn}) is conditioned on the classical
            information from the detectors), and decoder $A'$. In our
            implementation, the blocks $A$, $A'$, and $S_1$, which are
            composed of  two CNOT gates with two qubits in $| 0 \rangle$, can
            be replaced by type-II four-qubit mixing (4QM), i.e., by
            spontaneous evolution for the time $t=\pi/(2J^{(4)})$ of the
            system governed by the effective interaction Hamiltonian $\hat
            V^{(4)}$, given in Eq.~(\ref{V4}).} \label{ECC2}
    \end{figure*}

\section{How to apply qubit mixing in an error-correction code}
\label{ecc}

Here we present a possible application of the spontaneous time
evolution of the Dicke model, which leads to 4QM, for the qubit
encoding, decoding, and error-syndrome detection modules in an
ECC.

\subsection{Quantum repetition codes via qubit mixing}

As shown in Sec.~\ref{results}B, 4QM can occur in a system
described by the Dicke model in the ultrastrong-coupling regime,
which can be described by the effective Hamiltonian ($\hbar=1$)
 \beq
 \hat V_{\rm II}^{(4)} = J^{(4)}\, ( \hat \sigma_-^{(1)} \hat
 \sigma_+^{(2)} \hat \sigma_+^{(3)} \hat \sigma_+^{(4)} + {\rm
 H.c.})\, , \label{V4}
 \eeq
and referred to as type-II 4QM. The spontaneous evolution
operator $\hat U_t$ after the time $t'=\pi/(2J^{(4)})$ is simply
given by
\bea \hat U^{(4)}_{t'} = \exp\left(-i\hat V_{\rm
II}^{(4)} t'\right)=\hat I^{(4)} -\ket{0111}\bra{0111}
\nonumber \\
- \ket{1000}\bra{1000}- i(\ket{0111}\bra{1000}+{\rm
    H.c.}),
\label{U4} \eea
where $\hat I^{(4)}={\rm eye}(16)$ is the
four-qubit identity operator and, for simplicity, hereafter we
denote $\ket{0}\equiv\ket{g}$ and $\ket{1}\equiv\ket{e}$. Thus, an
initial arbitrary state $\ket{\psi_1}=a\ket{0}+b\ket{1}$ of a
single qubit and the other three qubits in the ground state
$\ket{0}$, is transformed as
 \beq
 \hat S_{4}\hat U^{(4)}_{t'}
 (a\ket{0}+b\ket{1})\ket{000} =\ket{0}(a\ket{000}+b\ket{111}),
 \label{U4a} \eeq
as shown in Fig.~\ref{RepetitionCodesFig}(c). Note that the
$(\pi/4)$-phase gate $\hat S_{n}={\rm diag}(1,i)$ can be applied
to any qubit $n$ except $n=1$. It is seen that the first qubit is
decoupled (disentangled) from the others at $t=0$ and $t=t'$, so
it can be discarded. Thus, Eq.~(\ref{U4}) effectively describes
the standard ($N=3$)-qubit repetition code
\bea \hat U^{(N)}_{\rm rep} &&\hspace*{-4mm} \ket{\psi_1}\ket{0}^{\otimes (N-1)}\nn\\
   &=&\hat U_{\rm CNOT}^{12}\hat U_{\rm CNOT}^{13}\cdots\hat U_{\rm
    CNOT}^{1N}\ket{\psi_1}\ket{0}^{\otimes (N-1)} \nn\\
   &=&a\ket{0}^{\otimes
    N}+b\ket{1}^{\otimes N}\equiv \ket{\psi_N},
\label{RepetitionCodeEq} \eea
where $\hat U_{\rm CNOT}^{1n}$
denotes the CNOT gate with the control (target) qubit 1 ($n$) for
$n=2,...,N$.  More precisely,
$\hat S_{4}\hat U_{t'}^{(4)}\ket{\psi_{1}}\ket{000}=(\hat I^{(1)}\otimes
\hat U^{(3)}_{\rm rep})\ket{0}\ket{\psi_{1}}\ket{00}$,
where $\hat I^{(1)}$ is the
single-qubit identity operator and we have an auxiliary qubit in
the state $\ket{0}$ in addition to those in
Eq.~(\ref{RepetitionCodeEq}). Analogously, it can be shown that
the 3QM in the model described by the effective interaction
Hamiltonian
 \beq
 \hat V^{(3)} = J^{(3)}\, (\hat \sigma_+^{(1)}
 \hat \sigma_+^{(2)} \hat \sigma_-^{(3)} + {\rm H.c.})\, ,
 \label{V3}
 \eeq
after the time $t'=\pi/(2J^{(3)})$ is given by
\bea
\hat U^{(3)}_{t'} &=& \exp\left(-i\hat V^{(3)} t'\right) =
\hat I^{(3)}  -\ket{011}\bra{011}-\ket{100}\bra{100} \nn\\
&& -i(\ket{011}\bra{100}+{\rm H.c.}),
\label{U3} \eea
which realizes the two-qubit repetition code, as
shown in Fig.~\ref{RepetitionCodesFig}(b), since
 \beq \hat
 S_{3}\hat U^{(3)}_{t'} (a\ket{0}+b\ket{1})\ket{00}
 =\ket{0}(a\ket{00}+b\ket{11}). \label{U3a}
 \eeq
In Eq.~(\ref{U3}),
$\hat I^{(3)}={\rm eye}(8)$ is the three-qubit identity operator.

\subsection{Error-correction code}

Now we present the implementation of a toy version of an ECC (for
a pedagogical description, see~\cite{PerryBook}) for correcting a
single phase-flip error or a bit-flip error. We apply the ECC
circuit shown in Fig.~\ref{ECC2}, and show that the quantum
repetition codes, based on the CNOT gates in the modules depicted
in yellow, can be replaced by the type-II 4QM described above.

Figure~\ref{RepetitionCodesFig}(a) shows a standard $N$-qubit
repetition code implemented with  $N$ CNOT gates. Although CNOT
gates have been implemented in circuit-QED systems, these
implementations are based on a sequence of a few pulses using,
usually, higher-excited levels~\cite{DiCarlo09,DiCarlo10}. Thus,
it is desirable to reduce the number of CNOT gates or other
entangling gates, which cannot be implemented easily like those
corresponding to the spontaneous evolution of a given system.

Figure~\ref{ECC2} shows a standard circuit implementing the ECC,
which enables the correction of a single phase-flip error, or of a
single bit-flip error if the blocks $B$ and $B'$ are omitted.
These blocks $B$ and $B$' are composed of three single-qubit
rotations $\hat Y(\pi/2)$ and $\hat Y(-\pi/2)$ about angles of
$\pm\pi/2$ around the $y$-axis, respectively. The $\hat
Y$-rotation about an arbitrary angle $\theta$ is defined by $\hat
Y(\theta )=[\cos \frac{\theta}{2}, - \sin \frac{\theta}{2};\sin
\frac{\theta}{2},\cos \frac{\theta}{2} ]$. Thus, the rotations
$\hat Y(\pm\pi/2)$ effectively realize Hadamard-like gates
corresponding to rotations of the basis states
$\{\ket{0},\ket{1}\}$ into (from)
$\{\ket{\pm}=(\ket{+}+\ket{-})/\sqrt{2}\}$ in $B$ ($B'$). The
modules $A$ and $B$ ($B'$ and $A'$) enable encoding (decoding) a
single qubit against a single phase-flip error, while the gates
$A$ ($A'$), without $B$ and $B'$, enable encoding (decoding) a
single qubit against a single bit-flip error. The module $E$ in
Fig.~\ref{ECC2} describes a dissipative evolution or a channel,
which introduces a single error (either a bit flip or a phase
flip) in one of the qubits. The error-syndrome detection is
performed by the four CNOT gates in the modules $S_1$ and $S_2$
using two additional qubits, which are then measured by two
detectors yielding the values $m,n\in\{0,1\}$. The error
correction is done in the module $C$ by rotating (flipping) a
proper qubit as conditioned on the measured values $m,n$:
\beq
\hat R_{mn} = \hat\sigma^{(n-m+2)}_{x}\approx \hat
Y_{n-m+2}(\pi), \label{Rmn} \eeq
where we define
$\hat\sigma_x^{(4)}\approx\hat Y_{4}(\pi)$ to be the identity
operator.

As marked in yellow in Fig.~\ref{ECC2}, the encoder $A$, decoder
$A'$, but also the error-syndrome module $S_1$, can be implemented
via the type-II 4QM, i.e., the spontaneous evolution of the system
described by Eq.~(\ref{U4}). In fact, the error-syndrome module
$S_2$ can also be expressed via the sequence of the type-II 4QM
operations $\hat U^{(4)}$ applied to different qubits, together
with single-qubit rotations. This possibility comes from the
fundamental theorem about the universality of an arbitrary
entangling gate (thus, including $\hat U^{(4)}$). However, we
focus here on simple and direct applications of $\hat U^{(4)}$.
Thus, a lengthy circuit implementing the module $S_2$ via $\hat
U^{(4)}$ is not presented here.

Finally, we make three remarks:

(1) In our proposal we need to apply the 4QM between different
qubits. Thus, the question arises whether one can efficiently tune
the ultrastrong coupling between a chosen fraction of a collection
of six qubits and a cavity field. For switching on and off the
coupling it is possible to change the transition energy of the
qubits involved. This can be easily done by applying, e.g., a flux
offset to the qubits.

(2) By using three times the 4QM operations instead of the six
CNOT gates, we need an extra qubit in addition to the five qubits
shown in Fig.~\ref{RepetitionCodesFig}(c).

(3) In a specific physical system, it might be easier to perform
3QM rather than 4QM. Then, the six CNOT gates in the modules $A$,
$A'$, and $S_1$, shown in Fig.~\ref{RepetitionCodesFig}(c), can be
replaced by the 3QM corresponding to the spontaneous evolution
$\hat U^{(3)}_{t'}$, given by Eq.~(\ref{U3}).


%

\end{document}